\documentclass[reqno,12pt,a4paper]{amsart}

\usepackage{mathrsfs}
\usepackage{amssymb}
\usepackage{amsfonts}
\usepackage{latexsym}
\usepackage{amsthm}

\usepackage{graphicx}
\def\lb{\label}

\newcommand{\er}[1]{\textrm{(\ref{#1})}}

\begin{document}


\renewcommand{\theequation}{\arabic{section}.\arabic{equation}}
\theoremstyle{plain}
\newtheorem{theorem}{\bf Theorem}[section]
\newtheorem{lemma}[theorem]{\bf Lemma}
\newtheorem{corollary}[theorem]{\bf Corollary}
\newtheorem{proposition}[theorem]{\bf Proposition}
\newtheorem{definition}[theorem]{\bf Definition}
\newtheorem{remark}[theorem]{\it Remark}

\def\a{\alpha}  \def\cA{{\mathcal A}}     \def\bA{{\bf A}}  \def\mA{{\mathscr A}}
\def\b{\beta}   \def\cB{{\mathcal B}}     \def\bB{{\bf B}}  \def\mB{{\mathscr B}}
\def\g{\gamma}  \def\cC{{\mathcal C}}     \def\bC{{\bf C}}  \def\mC{{\mathscr C}}
\def\G{\Gamma}  \def\cD{{\mathcal D}}     \def\bD{{\bf D}}  \def\mD{{\mathscr D}}
\def\d{\delta}  \def\cE{{\mathcal E}}     \def\bE{{\bf E}}  \def\mE{{\mathscr E}}
\def\D{\Delta}  \def\cF{{\mathcal F}}     \def\bF{{\bf F}}  \def\mF{{\mathscr F}}
\def\c{\chi}    \def\cG{{\mathcal G}}     \def\bG{{\bf G}}  \def\mG{{\mathscr G}}
\def\z{\zeta}   \def\cH{{\mathcal H}}     \def\bH{{\bf H}}  \def\mH{{\mathscr H}}
\def\e{\eta}    \def\cI{{\mathcal I}}     \def\bI{{\bf I}}  \def\mI{{\mathscr I}}
\def\p{\psi}    \def\cJ{{\mathcal J}}     \def\bJ{{\bf J}}  \def\mJ{{\mathscr J}}
\def\vT{\Theta} \def\cK{{\mathcal K}}     \def\bK{{\bf K}}  \def\mK{{\mathscr K}}
\def\k{\kappa}  \def\cL{{\mathcal L}}     \def\bL{{\bf L}}  \def\mL{{\mathscr L}}
\def\l{\lambda} \def\cM{{\mathcal M}}     \def\bM{{\bf M}}  \def\mM{{\mathscr M}}
\def\L{\Lambda} \def\cN{{\mathcal N}}     \def\bN{{\bf N}}  \def\mN{{\mathscr N}}
\def\m{\mu}     \def\cO{{\mathcal O}}     \def\bO{{\bf O}}  \def\mO{{\mathscr O}}
\def\n{\nu}     \def\cP{{\mathcal P}}     \def\bP{{\bf P}}  \def\mP{{\mathscr P}}
\def\r{\rho}    \def\cQ{{\mathcal Q}}     \def\bQ{{\bf Q}}  \def\mQ{{\mathscr Q}}
\def\s{\sigma}  \def\cR{{\mathcal R}}     \def\bR{{\bf R}}  \def\mR{{\mathscr R}}
                \def\cS{{\mathcal S}}     \def\bS{{\bf S}}  \def\mS{{\mathscr S}}
\def\t{\tau}    \def\cT{{\mathcal T}}     \def\bT{{\bf T}}  \def\mT{{\mathscr T}}
\def\f{\phi}    \def\cU{{\mathcal U}}     \def\bU{{\bf U}}  \def\mU{{\mathscr U}}
\def\F{\Phi}    \def\cV{{\mathcal V}}     \def\bV{{\bf V}}  \def\mV{{\mathscr V}}
\def\P{\Psi}    \def\cW{{\mathcal W}}     \def\bW{{\bf W}}  \def\mW{{\mathscr W}}
\def\o{\omega}  \def\cX{{\mathcal X}}     \def\bX{{\bf X}}  \def\mX{{\mathscr X}}
\def\x{\xi}     \def\cY{{\mathcal Y}}     \def\bY{{\bf Y}}  \def\mY{{\mathscr Y}}
\def\X{\Xi}     \def\cZ{{\mathcal Z}}     \def\bZ{{\bf Z}}  \def\mZ{{\mathscr Z}}
\def\O{\Omega}

\newcommand{\gA}{\mathfrak{A}}          \newcommand{\ga}{\mathfrak{a}}
\newcommand{\gB}{\mathfrak{B}}          \newcommand{\gb}{\mathfrak{b}}
\newcommand{\gC}{\mathfrak{C}}          \newcommand{\gc}{\mathfrak{c}}
\newcommand{\gD}{\mathfrak{D}}          \newcommand{\gd}{\mathfrak{d}}
\newcommand{\gE}{\mathfrak{E}}
\newcommand{\gF}{\mathfrak{F}}           \newcommand{\gf}{\mathfrak{f}}
\newcommand{\gG}{\mathfrak{G}}           
\newcommand{\gH}{\mathfrak{H}}           \newcommand{\gh}{\mathfrak{h}}
\newcommand{\gI}{\mathfrak{I}}           \newcommand{\gi}{\mathfrak{i}}
\newcommand{\gJ}{\mathfrak{J}}           \newcommand{\gj}{\mathfrak{j}}
\newcommand{\gK}{\mathfrak{K}}            \newcommand{\gk}{\mathfrak{k}}
\newcommand{\gL}{\mathfrak{L}}            \newcommand{\gl}{\mathfrak{l}}
\newcommand{\gM}{\mathfrak{M}}            \newcommand{\gm}{\mathfrak{m}}
\newcommand{\gN}{\mathfrak{N}}            \newcommand{\gn}{\mathfrak{n}}
\newcommand{\gO}{\mathfrak{O}}
\newcommand{\gP}{\mathfrak{P}}             \newcommand{\gp}{\mathfrak{p}}
\newcommand{\gQ}{\mathfrak{Q}}             \newcommand{\gq}{\mathfrak{q}}
\newcommand{\gR}{\mathfrak{R}}             \newcommand{\gr}{\mathfrak{r}}
\newcommand{\gS}{\mathfrak{S}}              \newcommand{\gs}{\mathfrak{s}}
\newcommand{\gT}{\mathfrak{T}}             \newcommand{\gt}{\mathfrak{t}}
\newcommand{\gU}{\mathfrak{U}}             \newcommand{\gu}{\mathfrak{u}}
\newcommand{\gV}{\mathfrak{V}}             \newcommand{\gv}{\mathfrak{v}}
\newcommand{\gW}{\mathfrak{W}}             \newcommand{\gw}{\mathfrak{w}}
\newcommand{\gX}{\mathfrak{X}}               \newcommand{\gx}{\mathfrak{x}}
\newcommand{\gY}{\mathfrak{Y}}              \newcommand{\gy}{\mathfrak{y}}
\newcommand{\gZ}{\mathfrak{Z}}             \newcommand{\gz}{\mathfrak{z}}

\def\ve{\varepsilon} \def\vt{\vartheta} \def\vp{\varphi}  \def\vk{\varkappa}

\def\Z{{\mathbb Z}} \def\R{{\mathbb R}} \def\C{{\mathbb C}}  \def\K{{\mathbb K}}
\def\T{{\mathbb T}} \def\N{{\mathbb N}} \def\dD{{\mathbb D}} \def\S{{\mathbb S}}
\def\B{{\mathbb B}}


\def\la{\leftarrow}              \def\ra{\rightarrow}     \def\Ra{\Rightarrow}
\def\ua{\uparrow}                \def\da{\downarrow}
\def\lra{\leftrightarrow}        \def\Lra{\Leftrightarrow}
\newcommand{\abs}[1]{\lvert#1\rvert}
\newcommand{\br}[1]{\left(#1\right)}

\def\lan{\langle} \def\ran{\rangle}


\def\lt{\biggl}                  \def\rt{\biggr}
\def\ol{\overline}               \def\wt{\widetilde}
\def\no{\noindent}


\let\ge\geqslant                 \let\le\leqslant
\def\lan{\langle}                \def\ran{\rangle}
\def\/{\over}                    \def\iy{\infty}
\def\sm{\setminus}               \def\es{\emptyset}
\def\ss{\subset}                 \def\ts{\times}
\def\pa{\partial}                \def\os{\oplus}
\def\om{\ominus}                 \def\ev{\equiv}
\def\iint{\int\!\!\!\int}        \def\iintt{\mathop{\int\!\!\int\!\!\dots\!\!\int}\limits}
\def\el2{\ell^{\,2}}             \def\1{1\!\!1}
\def\sh{\sharp}
\def\wh{\widehat}
\def\bs{\backslash}
\def\na{\nabla}

\def\ch{\mathop{\mathrm{ch}}\nolimits}
\def\sh{\mathop{\mathrm{sh}}\nolimits}
\def\all{\mathop{\mathrm{all}}\nolimits}
\def\Area{\mathop{\mathrm{Area}}\nolimits}
\def\arg{\mathop{\mathrm{arg}}\nolimits}
\def\const{\mathop{\mathrm{const}}\nolimits}
\def\det{\mathop{\mathrm{det}}\nolimits}
\def\diag{\mathop{\mathrm{diag}}\nolimits}
\def\diam{\mathop{\mathrm{diam}}\nolimits}
\def\dim{\mathop{\mathrm{dim}}\nolimits}
\def\dist{\mathop{\mathrm{dist}}\nolimits}
\def\Im{\mathop{\mathrm{Im}}\nolimits}
\def\Iso{\mathop{\mathrm{Iso}}\nolimits}
\def\Ker{\mathop{\mathrm{Ker}}\nolimits}
\def\Lip{\mathop{\mathrm{Lip}}\nolimits}
\def\rank{\mathop{\mathrm{rank}}\limits}
\def\Ran{\mathop{\mathrm{Ran}}\nolimits}
\def\Re{\mathop{\mathrm{Re}}\nolimits}
\def\Res{\mathop{\mathrm{Res}}\nolimits}
\def\res{\mathop{\mathrm{res}}\limits}
\def\sign{\mathop{\mathrm{sign}}\nolimits}
\def\span{\mathop{\mathrm{span}}\nolimits}
\def\supp{\mathop{\mathrm{supp}}\nolimits}
\def\Tr{\mathop{\mathrm{Tr}}\nolimits}
\def\BBox{\hspace{1mm}\vrule height6pt width5.5pt depth0pt \hspace{6pt}}
\def\where{\mathop{\mathrm{where}}\nolimits}
\def\as{\mathop{\mathrm{as}}\nolimits}
\def\Dom{\mathop{\mathrm{Dom}}\nolimits}


\newcommand\nh[2]{\widehat{#1}\vphantom{#1}^{(#2)}}
\def\dia{\diamond}

\def\Oplus{\bigoplus\nolimits}



\def\qqq{\qquad}
\def\qq{\quad}
\let\ge\geqslant
\let\le\leqslant
\let\geq\geqslant
\let\leq\leqslant
\newcommand{\ca}{\begin{cases}}
\newcommand{\ac}{\end{cases}}
\newcommand{\ma}{\begin{pmatrix}}
\newcommand{\am}{\end{pmatrix}}
\renewcommand{\[}{\begin{equation}}
\renewcommand{\]}{\end{equation}}
\def\eq{\begin{equation}}
\def\qe{\end{equation}}
\def\[{\begin{equation}}
\def\bu{\bullet}

\newcommand{\fr}{\frac}
\newcommand{\tf}{\tfrac}

\title[{Resonances  of 4-th order differential operators}]
{Resonances of 4-th Order Differential Operators}

\date{\today}
\author[Andrey Badanin]{Andrey Badanin}
\author[Evgeny Korotyaev]{Evgeny L. Korotyaev}
\address{Saint-Petersburg
State University, Universitetskaya nab. 7/9, St. Petersburg, 199034
Russia,
an.badanin@gmail.com,\  a.badanin@spbu.ru,\
korotyaev@gmail.com,\  e.korotyaev@spbu.ru}


\begin{abstract}
\no We consider fourth order ordinary differential operator with
compactly supported  coefficients  on the line. We define resonances
as zeros of the Fredholm determinant which is analytic on a four
sheeted Riemann surface. We determine asymptotics of the number of
resonances in complex discs at large radius. We consider resonances
of an Euler-Bernoulli operator on the real line with the positive
coefficients which  are constants outside some finite interval. We
show that the Euler-Bernoulli operator has no eigenvalues and
resonances iff the positive coefficients are constants on the whole
axis.

\end{abstract}

\maketitle

\section {Introduction and main results}
\setcounter{equation}{0}

\subsection{Introduction}
 There are a lot
of results about resonances for 1-dim second order operators,  see
e.g., \cite{F97}, \cite{H99}, \cite{K04}, \cite{S00},  \cite{Z87}
and references therein. We can say that problems of resonances for
these operators are well understood. The resonance scattering for
third order operators on the line was considered in \cite{K16}.
There are a lot of papers \cite{B85}, \cite{I88}, \cite{Iw88}, ....
and even a book \cite{BDT88} about scattering for one dimensional
higher order $\ge 3$ operators. Unfortunately, even the inverse
scattering problems for higher order operators on the line are not
solved and there are few  results about resonances \cite{K16},
\cite{BK17}.

We discuss resonances of fourth order  differential operators $H$
with compactly supported  coefficients  on the line given by
\[
\label{a.1}
H=H_0+V, \qqq V=2\pa p\pa +q,\qqq \pa ={d\/dx}.
\]
where the operator $H_0=\pa^4$ is  unperturbed.
Below we assume that our coefficients $p,q\in \cH_0$, where $\cH_m,m=0,1,2,...$,
are the spaces of compactly supported functions
$$
\cH_m=\cH_m(\g)=\big\{f\in L^1(\R): \supp f\ss[0,\g],f^{(m)}\in
L^1(0,\g)\big\}
$$
for some $\g>0$. Our operator $H$
is self-adjoint on the corresponding form domain (see Sect.~2).
The operator $H$ has
purely absolutely continuous spectrum $[0,\iy)$ plus a finite number
 of simple eigenvalues on the real line, see Theorem~\ref{T1}.

We define the Fredholm determinant by \er{a.2}, which is  analytic
on a four sheeted Riemann surface of the function $\l^{1\/4}$. We
define a resonance  of the operator $H$ as a zero of the Fredholm
determinant  on the non-physical sheets of this surface.

Our main goal is to obtain estimates of $D$ and determine
asymptotics of the number of resonances in the large disc. Moreover,
we derive trace formulas in terms of resonances and prove a Borg
type results about resonances for Euler-Bernoulli operator.

\subsection{Schr\"odinger operator}
In  order to discuss resonances for fourth order operators we
consider a Schr\"odinger operator $h$ on $L^2(\R)$ given by
$$
h=h_0-p,\qqq h_0=-{d^2\/dx^2},
$$
 where $h_0$ is the unperturbed operator and the compactly
supported  potential  $p\in\cH_0$.
 We recall the well known results for the operator $h$, see, e.g.,
\cite{DT79}, \cite{Fa64}. The operator $h$ has purely absolutely
continuous spectrum $[0,\iy)$ plus a finite number of simple
eigenvalues $e_1<e_2<...<e_N<0$.

We define the resolvent $r_0(k)$ and the operator $ y_0(k)$ for
$k\in \C_+$  by
$$
r_0(k)=(h_0-k^2)^{-1},\qqq  y_0(k)=|p|^{1\/2}r_0(k)p^{1\/2}, \qq
\where \qq p^{1\/2}=|p|^{1\/2}\sign p.
$$
 Each operator $y_0(k), k\in \C_+$, is trace class and thus we can introduce
the Fredholm determinant by
\[
\lb{defd}
d(k)=\det (I+y_0(k)),\qqq k\in \C_+.
\]
The function $d(k)$ is analytic in $ \C_+$ and has an analytic
extension onto the whole complex plane without zero. It has simple
zeros (the eigenvalues) $i|e_1|^{1\/2},..., i|e_N|^{1\/2}\in i\R_+$
in $\C_+$ and there are no other zeros in $\C_+$. Moreover, $d(k)$
has an infinite number of zeros (resonances) in $\C_-$. The operator
$h$ is self-adjoint and then the set of resonances is symmetric with
respect to the imaginary axis, see~Fig.\ref{FigRSchr}.

\begin{figure}[t]
\tiny
\unitlength 0.7mm
\linethickness{0.4pt}
\ifx\plotpoint\undefined\newsavebox{\plotpoint}\fi 
\begin{picture}(109.25,94.25)(0,0)
\put(4.5,48.25){\line(1,0){104.75}}
\put(54.5,94.25){\line(0,-1){88.5}}
\put(75.75,40.25){\circle*{1.75}}
\put(33.25,40.25){\circle*{1.75}}
\put(86.5,33.75){\circle*{1.75}}
\put(22.5,33.75){\circle*{1.75}}
\put(88,40.75){\circle*{1.75}}
\put(21,40.75){\circle*{1.75}}
\put(95,33.5){\circle*{1.75}}
\put(14,33.5){\circle*{1.75}}
\put(54.5,25){\circle*{1.75}}
\put(54.5,19.5){\circle*{1.75}}
\put(54.5,28.5){\circle*{1.75}}
\put(54.5,11.25){\circle*{1.75}}
\put(60.75,42.25){\circle*{1.75}}
\put(48.25,42.25){\circle*{1.75}}
\put(60.75,35.25){\circle*{1.75}}
\put(48.25,35.25){\circle*{1.75}}
\put(83.75,44.25){\circle*{1.75}}
\put(25.25,44.25){\circle*{1.75}}
\put(103,33.5){\circle*{1.75}}
\put(6,33.5){\circle*{1.75}}
\put(107.75,31.5){\circle*{1.75}}
\put(1.25,31.5){\circle*{1.75}}
\put(71,46){\circle*{1.75}}
\put(38,46){\circle*{1.75}}
\put(54.5,67){\circle*{1.75}}
\put(54.5,83.5){\circle*{1.75}}
\put(54.5,92.5){\circle*{1.75}}
\put(108.75,50.5){\makebox(0,0)[cc]{$\Re k$}}
\put(48.25,90){\makebox(0,0)[cc]{$\Im k$}}
\qbezier(23.75,48)(23.75,42.5)(3.75,38)
\qbezier(85.25,48)(85.25,42.5)(105.25,38)
\put(61.75,67.5){\makebox(0,0)[cc]{$i|e_1|^{1\/2}$}}
\put(61.75,84.25){\makebox(0,0)[cc]{$i|e_2|^{1\/2}$}}
\put(61.75,93.5){\makebox(0,0)[cc]{$i|e_3|^{1\/2}$}}
\qbezier(95.25,51.75)(95.625,44)(94.5,38.25)
\qbezier(13.75,51.75)(13.375,44)(14.5,38.25)
\put(95.5,43.25){\line(4,5){4.05}}
\put(98.0,40){\line(4,5){6.5}}
\put(102.25,39){\line(4,5){7.0}}
\put(13.5,45){\line(-3,-4){4.05}}
\put(10.75,48){\line(-3,-4){7.0}}
\put(5.5,48){\line(-3,-4){6}}
\end{picture}
\caption{\footnotesize
Resonances for the Schr\"odinger operator $h$.
The resonances are marked by circles.
The forbidden domain for the resonances is shaded.}
\lb{FigRSchr}
\end{figure}
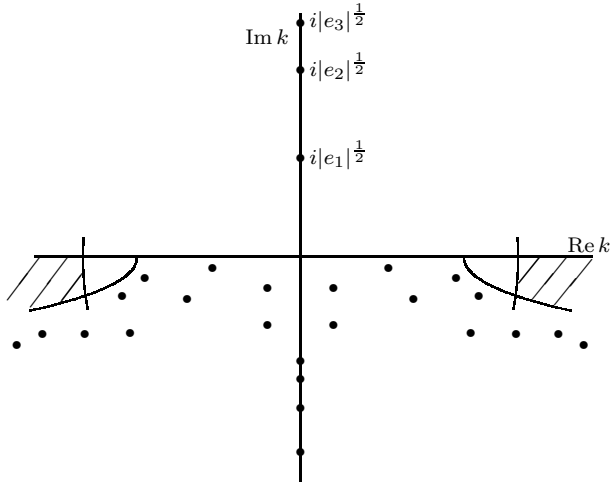

The problems of resonances for 1-dimensional Schr\"odinger
operators with compactly supported potentials are well understood.
Recall the following results:

$\bu$  Let  $\g=\sup (\supp(p))$ and let  $n(r)$ be the number  of
zeros of $d(k)$ in a disk $|k|<r$. Zworski \cite{Z87} determined the
following asymptotics (see also \cite{F97}, \cite{K05}, \cite{S00})
$$
n(r)={2\g r\/\pi}+o(r)\qqq\as\qq r\to\iy.
$$
For each $\d >0$ the number of zeros of $d(k)$ with
modulus $\leq r$ lying outside both of the two sectors $|\arg k| ,
|\arg k -\pi |<\d$ is $o(r)$ for large $r$.

$\bu$  There are only finitely many resonances in the domain $\{k\in
\C_-:|k|>\|p\|e^{\|p\|}e^{-2\g\Im k}\}$.

$\bu$  The resonances may have any multiplicity (see \cite{K05}).

$\bu$  Inverse resonance  problem (uniqueness, characterization and
recovering) was solved in terms of resonances for the Schr\"odinger
operator with a compactly supported potential on the real line
\cite{K05}.

$\bu$  Stability estimates for resonances were determined in
\cite{K04x} and \cite{MSW10}.

$\bu$  Lieb-Thirring type inequalities for resonances were obtained
 in \cite{K16xx}.

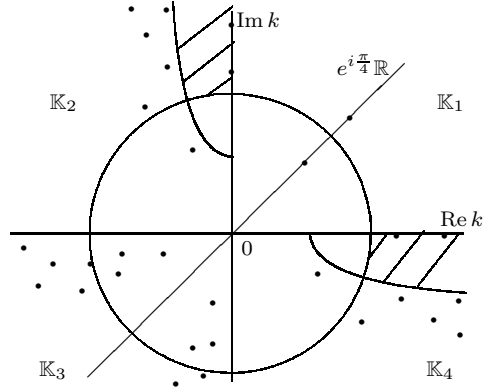
\begin{figure}[t]
\tiny
\unitlength 0.7mm
\linethickness{0.4pt}
\ifx\plotpoint\undefined\newsavebox{\plotpoint}\fi 
\begin{picture}(89.75,78.5)(0,0)
\put(45.5,76.25){\line(0,-1){70}}
\put(89,34.25){\line(-1,0){85}}
\put(18.5,7){\line(1,1){59.5}}
\put(86.75,59){\makebox(0,0)[cc]{$\K_1$}}
\put(14,59){\makebox(0,0)[cc]{$\K_2$}}
\put(12,8.25){\makebox(0,0)[cc]{$\K_3$}}
\put(84.75,8.25){\makebox(0,0)[cc]{$\K_4$}}
\put(70.25,66.25){\makebox(0,0)[cc]{$e^{i{\pi\/4}}\R$}}
\put(48.5,31.5){\makebox(0,0)[cc]{$0$}}
\put(88.75,37){\makebox(0,0)[cc]{$\Re k$}}
\put(50.5,74.75){\makebox(0,0)[cc]{$\Im k$}}
\qbezier(60.25,34.25)(60.625,24.875)(89.5,23)
\qbezier(45.75,48.75)(36.375,49.125)(34.5,78)
\multiput(36.5,63)(.042600897,.033632287){223}{\line(1,0){.042600897}}
\multiput(74.25,25.25)(.033632287,.042600897){223}{\line(0,1){.042600897}}
\multiput(35.5,69.25)(.043067227,.033613445){238}{\line(1,0){.043067227}}
\multiput(80.5,24.25)(.033613445,.043067227){238}{\line(0,1){.043067227}}
\put(34.5,75.5){\line(4,3){4}}
\put(86.75,23.25){\line(3,4){3}}
\put(71.69,34.25){\line(0,1){1.1155}}
\put(71.666,35.366){\line(0,1){1.1135}}
\put(71.596,36.479){\line(0,1){1.1096}}
\multiput(71.478,37.589)(-.032842,.220723){5}{\line(0,1){.220723}}
\multiput(71.314,38.692)(-.03009,.156529){7}{\line(0,1){.156529}}
\multiput(71.104,39.788)(-.032084,.13573){8}{\line(0,1){.13573}}
\multiput(70.847,40.874)(-.033584,.119338){9}{\line(0,1){.119338}}
\multiput(70.545,41.948)(-.031572,.096394){11}{\line(0,1){.096394}}
\multiput(70.197,43.008)(-.032644,.087062){12}{\line(0,1){.087062}}
\multiput(69.806,44.053)(-.0334964,.0790218){13}{\line(0,1){.0790218}}
\multiput(69.37,45.08)(-.0318938,.0671998){15}{\line(0,1){.0671998}}
\multiput(68.892,46.088)(-.0325318,.0616822){16}{\line(0,1){.0616822}}
\multiput(68.371,47.075)(-.0330403,.0567103){17}{\line(0,1){.0567103}}
\multiput(67.81,48.039)(-.0334366,.0521955){18}{\line(0,1){.0521955}}
\multiput(67.208,48.979)(-.0337348,.0480679){19}{\line(0,1){.0480679}}
\multiput(66.567,49.892)(-.0323297,.0421636){21}{\line(0,1){.0421636}}
\multiput(65.888,50.777)(-.0325307,.0389092){22}{\line(0,1){.0389092}}
\multiput(65.172,51.633)(-.0326589,.0358715){23}{\line(0,1){.0358715}}
\multiput(64.421,52.458)(-.0327206,.0330258){24}{\line(0,1){.0330258}}
\multiput(63.636,53.251)(-.0355668,.0329905){23}{\line(-1,0){.0355668}}
\multiput(62.818,54.01)(-.0386055,.0328906){22}{\line(-1,0){.0386055}}
\multiput(61.968,54.733)(-.0418616,.0327198){21}{\line(-1,0){.0418616}}
\multiput(61.089,55.421)(-.045365,.0324707){20}{\line(-1,0){.045365}}
\multiput(60.182,56.07)(-.0491522,.0321345){19}{\line(-1,0){.0491522}}
\multiput(59.248,56.681)(-.0564011,.0335654){17}{\line(-1,0){.0564011}}
\multiput(58.289,57.251)(-.0613775,.0331031){16}{\line(-1,0){.0613775}}
\multiput(57.307,57.781)(-.0669008,.0325164){15}{\line(-1,0){.0669008}}
\multiput(56.304,58.269)(-.0730855,.0317838){14}{\line(-1,0){.0730855}}
\multiput(55.28,58.714)(-.086755,.033451){12}{\line(-1,0){.086755}}
\multiput(54.239,59.115)(-.096097,.032466){11}{\line(-1,0){.096097}}
\multiput(53.182,59.472)(-.107119,.031221){10}{\line(-1,0){.107119}}
\multiput(52.111,59.784)(-.135427,.033342){8}{\line(-1,0){.135427}}
\multiput(51.028,60.051)(-.156243,.031542){7}{\line(-1,0){.156243}}
\multiput(49.934,60.272)(-.183674,.029075){6}{\line(-1,0){.183674}}
\multiput(48.832,60.446)(-.27711,.03195){4}{\line(-1,0){.27711}}
\put(47.724,60.574){\line(-1,0){1.1128}}
\put(46.611,60.655){\line(-1,0){1.1152}}
\put(45.495,60.689){\line(-1,0){1.1157}}
\put(44.38,60.676){\line(-1,0){1.1141}}
\put(43.266,60.615){\line(-1,0){1.1106}}
\multiput(42.155,60.508)(-.221018,-.030792){5}{\line(-1,0){.221018}}
\multiput(41.05,60.354)(-.182935,-.033408){6}{\line(-1,0){.182935}}
\multiput(39.952,60.154)(-.136022,-.030822){8}{\line(-1,0){.136022}}
\multiput(38.864,59.907)(-.119645,-.032474){9}{\line(-1,0){.119645}}
\multiput(37.787,59.615)(-.096683,-.030676){11}{\line(-1,0){.096683}}
\multiput(36.724,59.278)(-.087361,-.031834){12}{\line(-1,0){.087361}}
\multiput(35.676,58.896)(-.0793294,-.0327613){13}{\line(-1,0){.0793294}}
\multiput(34.644,58.47)(-.0723139,-.033502){14}{\line(-1,0){.0723139}}
\multiput(33.632,58.001)(-.0619816,-.0319577){16}{\line(-1,0){.0619816}}
\multiput(32.64,57.489)(-.0570147,-.0325123){17}{\line(-1,0){.0570147}}
\multiput(31.671,56.937)(-.0525037,-.0329506){18}{\line(-1,0){.0525037}}
\multiput(30.726,56.343)(-.0483791,-.0332871){19}{\line(-1,0){.0483791}}
\multiput(29.807,55.711)(-.044585,-.0335337){20}{\line(-1,0){.044585}}
\multiput(28.915,55.04)(-.0410767,-.0336999){21}{\line(-1,0){.0410767}}
\multiput(28.052,54.333)(-.0361732,-.0323244){23}{\line(-1,0){.0361732}}
\multiput(27.22,53.589)(-.0333282,-.0324125){24}{\line(-1,0){.0333282}}
\multiput(26.42,52.811)(-.0333193,-.0352589){23}{\line(0,-1){.0352589}}
\multiput(25.654,52)(-.0332476,-.0382984){22}{\line(0,-1){.0382984}}
\multiput(24.923,51.158)(-.033107,-.041556){21}{\line(0,-1){.041556}}
\multiput(24.227,50.285)(-.0328905,-.0450616){20}{\line(0,-1){.0450616}}
\multiput(23.57,49.384)(-.0325895,-.0488517){19}{\line(0,-1){.0488517}}
\multiput(22.95,48.456)(-.0321938,-.0529711){18}{\line(0,-1){.0529711}}
\multiput(22.371,47.502)(-.0336716,-.0610675){16}{\line(0,-1){.0610675}}
\multiput(21.832,46.525)(-.0331361,-.066596){15}{\line(0,-1){.066596}}
\multiput(21.335,45.526)(-.032461,-.0727872){14}{\line(0,-1){.0727872}}
\multiput(20.881,44.507)(-.0316197,-.0797913){13}{\line(0,-1){.0797913}}
\multiput(20.47,43.47)(-.033357,-.095791){11}{\line(0,-1){.095791}}
\multiput(20.103,42.416)(-.032214,-.106825){10}{\line(0,-1){.106825}}
\multiput(19.781,41.348)(-.030754,-.120099){9}{\line(0,-1){.120099}}
\multiput(19.504,40.267)(-.032991,-.155943){7}{\line(0,-1){.155943}}
\multiput(19.273,39.175)(-.030779,-.183396){6}{\line(0,-1){.183396}}
\multiput(19.088,38.075)(-.027617,-.221437){5}{\line(0,-1){.221437}}
\put(18.95,36.968){\line(0,-1){1.112}}
\put(18.859,35.856){\line(0,-1){4.4568}}
\multiput(18.964,31.399)(.028738,-.221295){5}{\line(0,-1){.221295}}
\multiput(19.108,30.293)(.031708,-.183238){6}{\line(0,-1){.183238}}
\multiput(19.298,29.193)(.029558,-.136303){8}{\line(0,-1){.136303}}
\multiput(19.535,28.103)(.031362,-.119941){9}{\line(0,-1){.119941}}
\multiput(19.817,27.023)(.032755,-.10666){10}{\line(0,-1){.10666}}
\multiput(20.144,25.957)(.031021,-.087653){12}{\line(0,-1){.087653}}
\multiput(20.517,24.905)(.0320234,-.0796302){13}{\line(0,-1){.0796302}}
\multiput(20.933,23.87)(.0328292,-.0726219){14}{\line(0,-1){.0726219}}
\multiput(21.393,22.853)(.033473,-.0664273){15}{\line(0,-1){.0664273}}
\multiput(21.895,21.856)(.0319815,-.0573141){17}{\line(0,-1){.0573141}}
\multiput(22.438,20.882)(.0324617,-.0528074){18}{\line(0,-1){.0528074}}
\multiput(23.023,19.932)(.0328365,-.048686){19}{\line(0,-1){.048686}}
\multiput(23.647,19.007)(.0331183,-.0448944){20}{\line(0,-1){.0448944}}
\multiput(24.309,18.109)(.033317,-.0413878){21}{\line(0,-1){.0413878}}
\multiput(25.009,17.24)(.0334411,-.0381296){22}{\line(0,-1){.0381296}}
\multiput(25.744,16.401)(.0334974,-.0350897){23}{\line(0,-1){.0350897}}
\multiput(26.515,15.594)(.0349481,-.0336452){23}{\line(1,0){.0349481}}
\multiput(27.318,14.82)(.0379881,-.0336017){22}{\line(1,0){.0379881}}
\multiput(28.154,14.081)(.0412468,-.0334914){21}{\line(1,0){.0412468}}
\multiput(29.02,13.377)(.0447543,-.0333074){20}{\line(1,0){.0447543}}
\multiput(29.915,12.711)(.048547,-.0330417){19}{\line(1,0){.048547}}
\multiput(30.838,12.083)(.0526699,-.0326842){18}{\line(1,0){.0526699}}
\multiput(31.786,11.495)(.0571786,-.0322231){17}{\line(1,0){.0571786}}
\multiput(32.758,10.947)(.0621426,-.0316434){16}{\line(1,0){.0621426}}
\multiput(33.752,10.441)(.0724827,-.0331353){14}{\line(1,0){.0724827}}
\multiput(34.767,9.977)(.0794943,-.0323591){13}{\line(1,0){.0794943}}
\multiput(35.8,9.556)(.087521,-.031391){12}{\line(1,0){.087521}}
\multiput(36.851,9.18)(.106521,-.033205){10}{\line(1,0){.106521}}
\multiput(37.916,8.848)(.119808,-.031868){9}{\line(1,0){.119808}}
\multiput(38.994,8.561)(.136177,-.030133){8}{\line(1,0){.136177}}
\multiput(40.084,8.32)(.183102,-.032481){6}{\line(1,0){.183102}}
\multiput(41.182,8.125)(.221172,-.029672){5}{\line(1,0){.221172}}
\put(42.288,7.976){\line(1,0){1.1111}}
\put(43.399,7.875){\line(1,0){1.1144}}
\put(44.514,7.82){\line(1,0){1.1157}}
\put(45.629,7.813){\line(1,0){1.1151}}
\put(46.744,7.852){\line(1,0){1.1124}}
\multiput(47.857,7.939)(.27694,.03335){4}{\line(1,0){.27694}}
\multiput(48.965,8.072)(.183524,.030005){6}{\line(1,0){.183524}}
\multiput(50.066,8.252)(.156081,.032333){7}{\line(1,0){.156081}}
\multiput(51.158,8.479)(.120227,.030247){9}{\line(1,0){.120227}}
\multiput(52.24,8.751)(.10696,.031763){10}{\line(1,0){.10696}}
\multiput(53.31,9.068)(.095931,.032952){11}{\line(1,0){.095931}}
\multiput(54.365,9.431)(.0799241,.0312826){13}{\line(1,0){.0799241}}
\multiput(55.404,9.838)(.0729236,.0321535){14}{\line(1,0){.0729236}}
\multiput(56.425,10.288)(.0667352,.0328548){15}{\line(1,0){.0667352}}
\multiput(57.426,10.781)(.0612091,.0334135){16}{\line(1,0){.0612091}}
\multiput(58.406,11.315)(.0531065,.03197){18}{\line(1,0){.0531065}}
\multiput(59.361,11.891)(.0489888,.032383){19}{\line(1,0){.0489888}}
\multiput(60.292,12.506)(.0452,.0327){20}{\line(1,0){.0452}}
\multiput(61.196,13.16)(.0416953,.0329313){21}{\line(1,0){.0416953}}
\multiput(62.072,13.851)(.0384384,.0330857){22}{\line(1,0){.0384384}}
\multiput(62.917,14.579)(.0353992,.0331702){23}{\line(1,0){.0353992}}
\multiput(63.732,15.342)(.0325529,.0331911){24}{\line(0,1){.0331911}}
\multiput(64.513,16.139)(.0324768,.0360365){23}{\line(0,1){.0360365}}
\multiput(65.26,16.968)(.0323332,.0390734){22}{\line(0,1){.0390734}}
\multiput(65.971,17.827)(.0337215,.0444431){20}{\line(0,1){.0444431}}
\multiput(66.646,18.716)(.033491,.0482381){19}{\line(0,1){.0482381}}
\multiput(67.282,19.633)(.0331718,.0523642){18}{\line(0,1){.0523642}}
\multiput(67.879,20.575)(.0327526,.0568769){17}{\line(0,1){.0568769}}
\multiput(68.436,21.542)(.032219,.0618462){16}{\line(0,1){.0618462}}
\multiput(68.951,22.532)(.0315531,.0673605){15}{\line(0,1){.0673605}}
\multiput(69.425,23.542)(.0330958,.0791904){13}{\line(0,1){.0791904}}
\multiput(69.855,24.572)(.032202,.087226){12}{\line(0,1){.087226}}
\multiput(70.241,25.618)(.031084,.096553){11}{\line(0,1){.096553}}
\multiput(70.583,26.68)(.032979,.119507){9}{\line(0,1){.119507}}
\multiput(70.88,27.756)(.031396,.135891){8}{\line(0,1){.135891}}
\multiput(71.131,28.843)(.029297,.15668){7}{\line(0,1){.15668}}
\multiput(71.336,29.94)(.031724,.220886){5}{\line(0,1){.220886}}
\put(71.495,31.044){\line(0,1){1.1101}}
\put(71.607,32.154){\line(0,1){2.0956}}
\multiput(40.75,60.25)(.045673077,.033653846){104}{\line(1,0){.045673077}}
\multiput(71.5,29.75)(.033653846,.045673077){104}{\line(0,1){.045673077}}
\put(45.5,64.75){\circle*{1.118}}
\put(76.5,33.75){\circle*{1.118}}
\put(45.5,73.75){\circle*{1.118}}
\put(85.5,33.75){\circle*{1.118}}
\put(29.5,71.75){\circle*{1.118}}
\put(83.5,17.75){\circle*{1.118}}
\put(33.5,65.75){\circle*{1.118}}
\put(77.5,21.75){\circle*{1.118}}
\put(33.25,76.5){\circle*{1.118}}
\put(88.25,21.5){\circle*{1.118}}
\put(29.25,58.25){\circle*{1.118}}
\put(70,17.5){\circle*{1.118}}
\put(38.25,50){\circle*{1.118}}
\put(61.75,26.5){\circle*{1.118}}
\put(59.25,47.5){\circle*{1.118}}
\put(67.75,56){\circle*{1.118}}
\put(26.75,76.75){\circle*{1.118}}
\put(88.5,15){\circle*{1.118}}
\put(25,30.25){\circle*{1.118}}
\put(42,13.25){\circle*{1.118}}
\put(32.75,30.25){\circle*{1.118}}
\put(42,21){\circle*{1.118}}
\put(17.5,23.25){\circle*{1.118}}
\put(35,5.75){\circle*{1.118}}
\put(12,28.5){\circle*{1.118}}
\put(24.25,26.5){\circle*{1.118}}
\put(38.25,12.5){\circle*{1.118}}
\put(9.25,24.25){\circle*{1.118}}
\put(6.5,31.5){\circle*{1.118}}
\put(19,28.5){\circle*{1.118}}
\put(40.25,7.25){\circle*{1.118}}
\end{picture}
\caption{\footnotesize The plane of variable $k$. The function
$D(k)$ is real on the line $e^{i{\pi\/4}}\R$.
The resonances are marked by circles.
The forbidden domain for the resonances is shaded.}
\lb{FigPk}
\end{figure}

\subsection{Determinant}
In order to define the Fredholm determinant for the operator $H$ we
need a factorization of the perturbation $V$ in the form
\[
\lb{defVj} V=V_1V_2,\qqq V_1=(\pa |2p|^{1\/2}, |q|^{1\/2}),\qqq
 V_2=\ma
(2p)^{1\/2}\pa
\\ q^{1\/2}\am.
\]
We introduce a new spectral variable $k\in \K_1$ by $\l=k^4$, where
 the spectral parameter  $\l$ belongs to the cut plane $\C\sm [0,\iy)$ (see Fig~\ref{FigPk}) and
 the sector $\K_1$ is given   by
$$
\textstyle \K_1=\Big\{k\in\C: \arg k\in (0,{\pi\/2})\Big\}.
$$
Introduce the free resolvent $R_0(k)$ and the operator $Y_0(k)$ by
\[
\lb{Y0}
 R_0(k)=(H_0-k^4)^{-1},\qqq Y_0(k)=V_2 R_0(k)V_1, \qqq
k \in \K_1.
\]
In Proposition~\ref{T1} we show  that each
operator $Y_0(k),k\in \K_1,$ is trace class. Thus we can define
the Fredholm determinant $D(k),
k\in \K_1$, by
\[
\label{a.2}
D(k)=\det (I +Y_0(k)).
\]
If $k\in\ol\K_1\sm\{0\}$ is a zero of the determinant $D$, then
$\l=k^4\in \R\sm\{0\}$ and $\l$ is an eigenvalue of the operator
$H$. We present preliminary results  about the determinant.

\begin{proposition}
\lb{T1}

Let $p,q\in\cH_0$. Then

i) Each operator $Y_0(k),k\in \K_1,$ is trace class and
the operator-valued
function $k^3Y_0(k)$ is entire in the trace norm.

ii) The Fredholm determinant $D(k)$  is analytic in $\K_1$ and has
an analytic extension into the whole plane without zero such that
the function $k^4D(k)$ is entire. In particular, the operator $H$
has a finite number of eigenvalues. Moreover, $D(k)$ is real on the
line $e^{i{\pi\/4}}\R$ and satisfies:
\[
\lb{syr} D(k)=\ol{D(i\ol k)}\qqq \forall\ k\ne 0,
\]
\[
\lb{asDK+}
D(k)=1-\fr{1+i}{2k}\int_\R p(x)dx+\fr{O(1)}{k^{2}}\qqq
\as \qqq |k| \to \infty, \qq  k\in\ol\K_1,
\]
uniformly in $\arg k\in[0,{\pi\/2}]$.

\end{proposition}

\no {\bf Remark.} The function $D(k)$ is symmetric with respect to
the line $e^{i{\pi\/4}}\R$. Thus it is sufficiently to analyze this
function in the half-plane $e^{i{\pi\/4}}\C_+$.

\medskip

The zeros of the function $D$ in $\C\sm\ol\K_1$ are
called {\it resonances} of $H$. Let $\cN(r)$ be the number of zeros
of the function $D$ in the disc $|k|<r$, counted with multiplicity.
Introduce the domains
$$
\K_j=i^{j-1}\K_1,\qqq j=1,2,3,4.
$$
Now we formulate our first main result.

\begin{theorem}
\lb{CorEstN} Let $p,q\in\cH_0$. Then the determinant $D(k)$ and the
counting function $\cN(r)$ satisfy
\[
\lb{unifestD}
|D(k)|\le Ce^{2\g( (\Re k)_-+(\Im
k)_-)},\qqq\forall\qq k\in\C,\qq |k|\ge 1,
\]
\[
\lb{estnr} \cN(r)\le \fr{4\g r}{\pi}\big(1+o(1)\big)\qqq\as\qq
r\to\iy,
\]
where $(a)_-={|a|-a\/2}$. Moreover, if $k_*\in\K_2$ is a resonance,
then
\[
\lb{estFD} |k_*|\le Ce^{-2\g\Re k_*}.
\]
Here $C=C(p,q)>0$ is some constant depending on $p,q$ only.
\end{theorem}

\no {\bf Remark.} 1)  The estimate \er{estFD}  gives that the domain
$\{k\in\K_2:|k|>Ce^{-2\g\Re k}\}$ is forbidden for
resonances in $\K_2$ and, due to the symmetry of the function
$D(k)$, the domain $\{k\in\K_4:|k|>Ce^{-2\g\Im k}\}$ is
forbidden for resonances in $\K_4$, see Fig.~\ref{FigPk}.
The proof repeats arguments from \cite{K04}.

2) Estimate \er{unifestD} is crucial to prove trace formula
\er{expF'}  in terms of resonances.

3) Due to \er{syr} the set of resonances is  symmetric with respect
to the line  $e^{i{\pi\/4}}$ (with the same multiplicity).

\medskip
We describe the main difference between the case of the second and
fourth order operators.

$\bu$ Recall that for Schr\"odinger operators  the Riemann surface
in terms of the momentum $k$ is the complex plane. The upper
half-plane corresponds to the physical sheet and the lower
half-plane corresponds to the non-physical sheet. In order to
determine asymptotics of the determinant $d(k)$ of the Schr\"odinger
operator in the complex plane it is sufficiently to obtain the
asymptotics of the determinant and the $2\ts 2$ scattering
matrix in $\C_+$. The Birman-Krein formula (see \er{SD2od})
gives the asymptotics of the determinant in $\C_-$.

$\bu$ In the case of fourth order operators the Riemann surface has
4 sheets and each quadrant $\K_j\ss \C, j=1,2,3,4$ of the variable
$k=\l^{1\/4}$ corresponds to the sheet $\L_j$ of the Riemann surface
$\l^{1\/4}$. Using  arguments similar to the second order case we obtain
asymptotics of the determinant in the domains $\K_1,\K_2$ only
(and, by the symmetry, in $\K_4$). In fact, if we have asymptotics
of the determinant $D(k)$ and asymptotics of the $2\ts 2$
scattering matrix $S(k)$ in $\K_1$, then using the identity
\er{cS(k)} we obtain the asymptotics of the determinant $D(k)$ in
$\K_2$. In order to obtain the asymptotics in the domain $\K_3$ we
use additional arguments from \cite{K16} (for third order
operators), more complicated than for the sectors $\K_1,\K_2$. We
need to introduce an additional $4\ts4$ matrix-valued function
$\O(k)$, which satisfies the identity \er{cT(k)}. Having asymptotics
of the determinant $D(k)$ and of this $4\ts 4$
matrix $\O(k)$ in $\K_1$ and using the identity \er{cT(k)} we obtain
the asymptotics of the determinant $D(k)$ in $\K_3$. Note that in order
to determine asymptotics of determinants for $N$ order operators we
need to introduce the corresponding $N\ts N$ matrix-valued
functions.

\subsection{Asymptotics of resonances}

Introduce the Cartwright class of functions.
Recall that an entire function $f(z)$ is said to be of exponential type if there
is a constant $A$ such that $|f(z)|\leq\const e^{A|z|}$ everywhere.
The infimum of the set of $A$ for which such inequality   holds is
called the type of $f$. For each exponential type function $f$ we
define the types $\r _{\pm}(f)$ in $\C_\pm $ by
$$
\r _{\pm}(f)\equiv \lim \sup_{y\to \iy} {\log |f(\pm iy)|\/y} .
$$
The function $f$ is said to  belong  to the Cartwright class
$\cC_{Cart}$ if $f$ is entire, of exponential type, and satisfies
the following conditions:
$$
\int_\R\frac{\log(1+|f(x)|)}{1+x^2}dx
<\infty,\qq\rho_+(f)=0,\qq\rho_-(f)=2\rho>0,
$$
for some $\rho>0$. We recall the Levinson Theorem  (see
\cite{Ko88}):

{\it  Let an entire function $f$ be in the Cartwright class $
\cC_{Cart}$. Let $\cN(r,f)$ be the number of zeros of the function
$f$ in the disc $|k|<r$, counted with multiplicity. Then}
\[
\lb{AsN} \cN (r,f)={2\r\/ \pi }r+o(r)\qqq \as \qqq  r\to \iy.
\]

Roughly speaking the Fredholm determinants
of second order operators on the real
line with compactly supported potential belong to the Cartwright
class:

$\bu$  Schr\"odinger operators with compactly supported potentials
\cite{Z87},

$\bu$ Schr\"odinger operators with periodic plus compactly supported
potentials \cite{K11},

$\bu$ Schr\"odinger operators with matrix-valued compactly supported
potentials \cite{N07},

$\bu$ Dirac operators with $2\ts2$ matrix-valued compactly supported
potentials \cite{IK14}.

In all these cases the Fredholm determinants are entire and belong
to the Cartwright class, and the corresponding Riemann surfaces are
two sheeted. Thus the Levinson Theorem describes the distribution of
resonances in the large disc.

We underline that the Fredholm determinants of the Stark operator
with compactly supported potential does not belong to the Cartwright
class, since the order of  determinants is ${3\/2}$, see \cite{K16x}.
In this paper we show that the determinant
for a fourth order operators
does not belong to the Cartwright class also.

We determine asymptotics of resonances for fourth order operators
under   the stronger conditions for the coefficient $p$ and the
standard one for $q$:
\[
\lb{ppm}
(p,q)\in\cH_1\ts\cH_0,\qq p_+p_-\ne0,\qq\where\qq
p_-=p(+0),\ \ p_+=p(\g-0).
\]
Introduce the model numbers $k_n^0, n\in \pm
\N$ (see Fig.~\ref{figas}) by
$$
\begin{aligned}
k_n^0={1\/\g}\Big(i\pi j_n-2\log {4\pi n\/\g
|p_+p_-|^{1\/4}}\Big),\qq k_{-n}^0=ik_n^0,
\qq n\in\N,
\end{aligned}
$$
where
$$
j_n=\ca n,& \text{if}\ p_+p_->0\\
n+{1\/2},& \text{if}\ p_+p_-<0\ac .
$$

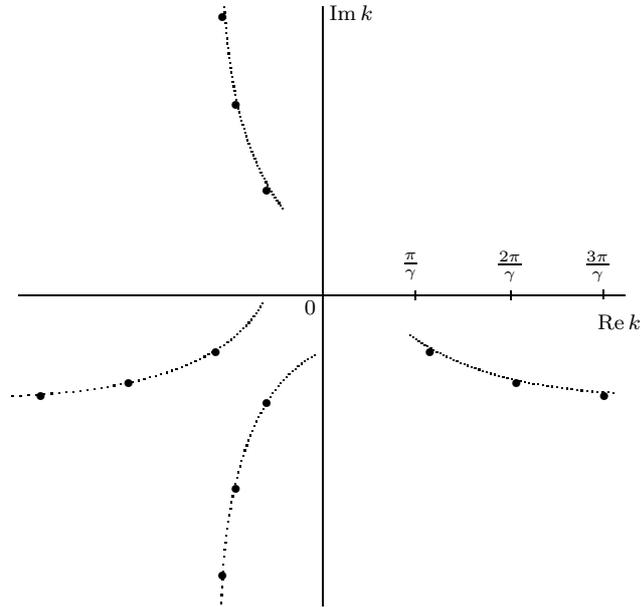
\begin{figure}[t]
\tiny
\unitlength 0.6mm 
\linethickness{0.4pt}
\ifx\plotpoint\undefined\newsavebox{\plotpoint}\fi 
\begin{picture}(137.25,137.75)(0,0)
\put(68.75,133.25){\line(0,-1){132.25}}
\put(2,69.5){\line(1,0){133}}
\put(92.25,57){\circle*{1.5}}
\put(56.5,92.75){\circle*{1.5}}
\put(45.25,57){\circle*{1.5}}
\put(56.5,45.75){\circle*{1.5}}
\put(111.25,50.25){\circle*{1.5}}
\put(49.75,111.75){\circle*{1.5}}
\put(26.25,50.25){\circle*{1.5}}
\put(49.75,26.75){\circle*{1.5}}
\put(130.5,47.25){\circle*{1.5}}
\put(46.75,131){\circle*{1.5}}
\put(7,47.25){\circle*{1.5}}
\put(46.75,7.5){\circle*{1.5}}
\put(89,70.5){\line(0,-1){2}}
\put(110,70.5){\line(0,-1){2}}
\put(130.25,70.75){\line(0,-1){2}}
\put(88,76){\makebox(0,0)[cc]{${\pi\/\g}$}}
\put(109.75,76){\makebox(0,0)[cc]{${2\pi\/\g}$}}
\put(129,76){\makebox(0,0)[cc]{${3\pi\/\g}$}}
\put(133.75,63.75){\makebox(0,0)[cc]{$\Re k$}}
\put(75,132){\makebox(0,0)[cc]{$\Im k$}}
\put(66,67){\makebox(0,0)[cc]{$0$}}
\bezier{50}(.75,47.25)(45.375,49)(55.5,67.75)
\bezier{50}(46.5,1.5)(48.25,46.125)(67,56.25)
\bezier{50}(60,88.75)(48.875,103.5)(47.25,133.25)
\bezier{50}(88,60.75)(102.75,49.625)(132.5,48)
\end{picture}
\caption{\footnotesize Asymptotics of resonances for step
coefficients} \lb{figas}
\end{figure}

\begin{theorem}
\lb{ThAsCF}

Let $p,q$ satisfy the conditions
\er{ppm}. Then for any $\ve\in (0,{\pi\/2\g})$ there exists $\r>0$
such that in each disk
$\{|k-k_{n}^0|<\ve\}\ss e^{i{\pi\/4}}\C_+\cap
\{|k|>\r\},n\in\pm\N$, there exists exactly one resonance $k_n$
 and there are no other
resonances in the domain $e^{i{\pi\/4}}\C_+\cap \{|k|>\r\}$.
These resonances satisfy
\[
\lb{asresiK+}
k_{n}=k_{n}^0+o(1)\qqq as\qq n\to \pm\iy.
\]
In particular, there are finitely many zeros of $D(k)$ on $\R\cup
i\R$. Moreover, let $\cN_j(r), j=2,3,4$ be  the number of zeros of
the function $D$ in a domain $\K_j\cap \{|k|<r\}$ counted with
multiplicity. Then
\[
\lb{asNj}
\begin{aligned}
\cN_2(r)=\cN_4(r)=\fr{\g r}{\pi}\big(1+o(1)\big),\\
\cN_3(r)=\fr{2\g r}{\pi}\big(1+o(1)\big)\qq
\end{aligned}
\]
as $r\to\iy$ and the determinant  $D$ does not belong to the
Cartwright class.

\end{theorem}

\no {\bf Remark.} From \er{asNj} we obtain
$\cN_3(r)=2\cN_2(r)(1+o(1))$. It means that the number of resonances
in the large disc in the domain $\K_3$ is in two times more than in
the domain $\K_2$, see Fig.~\ref{fighsq}.

\medskip

Our next results are devoted to trace formulas in terms of resonances.
 Recall that the function $D(k)$ may have a pole at the point $k=0$.
Then the function $D(k)$ satisfies
\[
\lb{Dat0} D(k)={\a\/ k^m}\big(1+\b k+O(k^2)\big)\qqq \as \qq |k|\to
0
\]
for some $\a,\b\in\C$, where $m\le 4$ is an order of the pole. Let
$\z_n,n\in\N$, be the zeros of the function $D$ in $\C\sm\{0\}$
labeled by $0<|\z_1|\le|\z_2|\le...$  counting with multiplicities.

\begin{theorem}
\lb{ThRes} Let $p,q\in\cH_0$ and let $R(k)=(H-k^4)^{-1}$. Then the
following trace formula
\[
\lb{expF'}
\Tr\big(R_0(k)-R(k)\big)={1\/4k^3}\Big(\b-{m\/k}+k\lim_{r\to\iy}\sum_{|\z_n|<r}
{1\/\z_n(k-\z_n)}\Big)
\]
holds true for all $k\in\C\sm\{0,\z_n,n\in \N\}$, where the series
converges absolutely and uniformly on any compact subset in
$\C\sm\{0,\z_n,n\in\N\}$, and $\b, m$ are defined in \er{Dat0}.
\end{theorem}

\no {\bf Remark.} 1) In the proof we use arguments from \cite{K04},
\cite{K16}.

2) A trace formula for the scattering phase function is proved in
Theorem~\ref{ThScATr}.

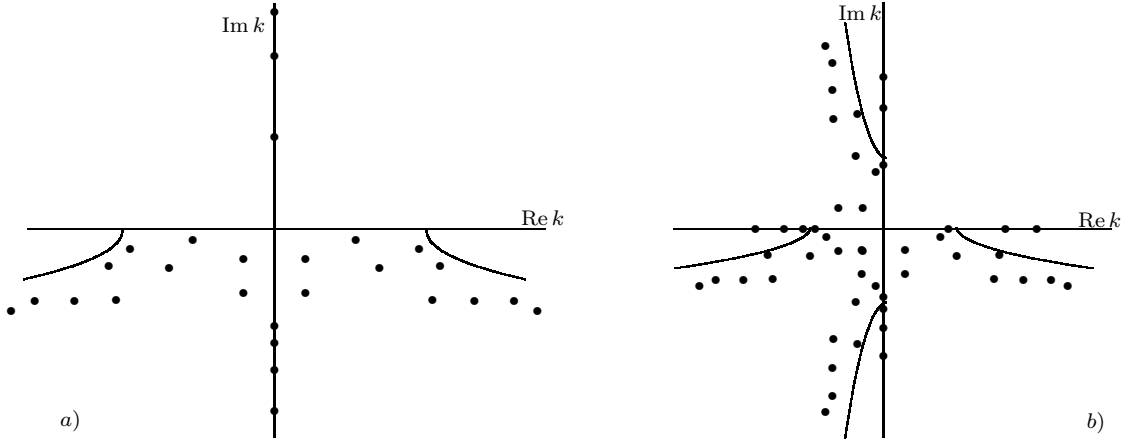
\begin{figure}[t]
\tiny
\unitlength 0.65mm
\linethickness{0.4pt}
\ifx\plotpoint\undefined
\newsavebox{\plotpoint}\fi 
\begin{picture}(223.796,94.428)(0,0)
\put(4.5,48.25){\line(1,0){104.75}}
\put(54.5,94.25){\line(0,-1){88.5}}
\put(177.75,94.428){\line(0,-1){88.589}}
\put(223.796,48.25){\line(-1,0){88.589}}
\put(13.25,9.25){\makebox(0,0)[cc]{$a)$}}
\put(220.799,8.843){\makebox(0,0)[cc]{$b)$}}
\put(75.75,40.25){\circle*{1.75}}
\put(192.625,42.65){\circle*{1.75}}
\put(172.15,33.375){\circle*{1.75}}
\put(33.25,40.25){\circle*{1.75}}
\put(162.875,42.65){\circle*{1.75}}
\put(172.15,63.125){\circle*{1.75}}
\put(86.5,33.75){\circle*{1.75}}
\put(200.15,38.1){\circle*{1.75}}
\put(167.6,25.85){\circle*{1.75}}
\put(22.5,33.75){\circle*{1.75}}
\put(155.35,38.1){\circle*{1.75}}
\put(167.6,70.65){\circle*{1.75}}
\put(88,40.75){\circle*{1.75}}
\put(201.2,43){\circle*{1.75}}
\put(172.5,24.8){\circle*{1.75}}
\put(21,40.75){\circle*{1.75}}
\put(154.3,43){\circle*{1.75}}
\put(172.5,71.7){\circle*{1.75}}
\put(95,33.5){\circle*{1.75}}
\put(206.1,37.925){\circle*{1.75}}
\put(167.425,19.9){\circle*{1.75}}
\put(14,33.5){\circle*{1.75}}
\put(149.4,37.925){\circle*{1.75}}
\put(167.425,76.6){\circle*{1.75}}
\put(54.5,25){\circle*{1.75}}
\put(177.75,31.975){\circle*{1.75}}
\put(161.475,48.25){\circle*{1.75}}
\put(54.5,19.5){\circle*{1.75}}
\put(177.75,28.125){\circle*{1.75}}
\put(157.625,48.25){\circle*{1.75}}
\put(54.5,28.5){\circle*{1.75}}
\put(177.75,34.425){\circle*{1.75}}
\put(163.925,48.25){\circle*{1.75}}
\put(54.5,11.25){\circle*{1.75}}
\put(177.75,22.35){\circle*{1.75}}
\put(151.85,48.25){\circle*{1.75}}
\put(60.75,42.25){\circle*{1.75}}
\put(182.125,44.05){\circle*{1.75}}
\put(173.55,43.875){\circle*{1.75}}
\put(48.25,42.25){\circle*{1.75}}
\put(173.375,44.05){\circle*{1.75}}
\put(173.55,52.625){\circle*{1.75}}
\put(60.75,35.25){\circle*{1.75}}
\put(182.125,39.15){\circle*{1.75}}
\put(168.65,43.875){\circle*{1.75}}
\put(48.25,35.25){\circle*{1.75}}
\put(173.375,39.15){\circle*{1.75}}
\put(168.65,52.625){\circle*{1.75}}
\put(83.75,44.25){\circle*{1.75}}
\put(25.25,44.25){\circle*{1.75}}
\put(103,33.5){\circle*{1.75}}
\put(211.7,37.925){\circle*{1.75}}
\put(167.425,14.3){\circle*{1.75}}
\put(6,33.5){\circle*{1.75}}
\put(143.8,37.925){\circle*{1.75}}
\put(167.425,82.2){\circle*{1.75}}
\put(107.75,31.5){\circle*{1.75}}
\put(215.025,36.525){\circle*{1.75}}
\put(166.025,10.975){\circle*{1.75}}
\put(1.25,31.5){\circle*{1.75}}
\put(140.475,36.525){\circle*{1.75}}
\put(166.025,85.525){\circle*{1.75}}
\put(71,46){\circle*{1.75}}
\put(189.3,46.675){\circle*{1.75}}
\put(176.175,36.7){\circle*{1.75}}
\put(38,46){\circle*{1.75}}
\put(166.2,46.675){\circle*{1.75}}
\put(176.175,59.8){\circle*{1.75}}
\put(54.5,67){\circle*{1.75}}
\put(177.75,61.375){\circle*{1.75}}
\put(190.875,48.25){\circle*{1.75}}
\put(54.5,83.5){\circle*{1.75}}
\put(177.75,72.925){\circle*{1.75}}
\put(202.425,48.25){\circle*{1.75}}
\put(54.5,92.5){\circle*{1.75}}
\put(177.75,79.225){\circle*{1.75}}
\put(208.725,48.25){\circle*{1.75}}
\put(108.75,50.5){\makebox(0,0)[cc]{$\Re k$}}
\put(221.5,50.25){\makebox(0,0)[cc]{$\Re k$}}
\put(48.25,90){\makebox(0,0)[cc]{$\Im k$}}
\put(173,92.5){\makebox(0,0)[cc]{$\Im k$}}
\qbezier(23.75,48)(23.75,42.5)(3.75,38)
\qbezier(85.25,48)(85.25,42.5)(105.25,38)
\qbezier(163,48.5)(162.25,44.125)(135.5,40.25)
\qbezier(178.25,62.75)(173.875,63.5)(170,90.25)
\qbezier(192.5,48.5)(193.25,44.125)(220,40.25)
\qbezier(178.25,33.25)(173.875,32.5)(170,5.75)
\end{picture}
\caption{\footnotesize a) Resonances for the second order operator
$h$; b) Resonances for the fourth order operator $h^2$. }
\lb{fighsq}
\end{figure}

\subsection{Euler-Bernoulli operators.}
We discuss  resonances of the Euler-Bernoulli operator
$$
\cE u={1\/b}(au'')'',
$$
acting on $L^2(\R,b(x)dx)$. We assume that the coefficients $a,b$
are positive, $a=b=1$ outside a unit interval and satisfy $
a-1,b-1\in\cH_4(1)$. The Euler-Bernoulli operator describes the
relationship between the thin beam's deflection and the applied
load, $a$ is the rigidity and $b$ is the density of the beam, see,
e.g., \cite{TW59}.

In Sect \ref{sect55} we show that
the operator $\cE\ge 0$ is unitarily equivalent to an
operator $H\ge 0$ with specific coefficients $p,q\in \cH_0$.
Then we can define a determinant
for the operator $\cE$ as the determinant
for the operator $H$ with these $p,q$.
Applying the results for the operator $H$
to the operator $\cE$ we obtain the following corollary
of Theorem~\ref{CorEstN}.

\begin{corollary}
\lb{CorEB} Let $a-1$ and $b-1\in\cH_4(1)$ and let $a,b$ be positive.
Then the determinant $D(k)$, the counting function $\cN(r)$ and the
resonances for the Euler-Bernoulli operator $\cE$ satisfy the
estimates \er{unifestD}--\er{estFD}, where
\[
\lb{gammaEB}
\g=\int_0^1\Big({b(x)\/a(x)}\Big)^{1\/4}dx.
\]
\end{corollary}

Now we formulate our result about  the inverse resonance scattering
for the Euler-Bernoulli operator $\cE$.

\begin{theorem}
\lb{ThEB} Let $a-1,b-1\in\cH_4$ and let $a,b$ be positive. Then the
operator $\cE$ does not have any eigenvalues and resonances iff $a=b=1$ on
the whole line.
\end{theorem}

\subsection{Historical review}
 There are a lot of results about resonances. We recall that
resonances, from a physicists point of view, were first studied by
Gamov \cite{Ga28}. Since then, properties of  resonances have been the
object of intense study and we refer to   \cite{SZ91} for the
mathematical approach in the multi-dimensional case  and references
given therein. We discuss the one-dimensional case. A lot of papers
are devoted to the resonances the one-dimensional Schr\"odinger
operators with compactly supported potentials, see Froese
\cite{F97}, Korotyaev \cite{K04}, Simon \cite{S00}, Zworski
\cite{Z87} and references given there.
 We recall that Zworski
\cite{Z87} obtained the first results about the asymptotic
distribution of resonances for the Schr\"odinger operator with
compactly supported potentials on the real line.
Inverse problems (characterization, recovering,
plus uniqueness) in terms of resonances were solved by Korotyaev for
the Schr\"odinger operator with a compactly supported potential on
the real line \cite{K05} and the half-line \cite{K04},
see also Zworski \cite{Z02}  concerning the uniqueness.

There are few papers devoted to systems. Nedelec \cite{N07}
considered resonances for Schr\"o\-dinger operators with compactly
supported matrix-valued potentials on the real line.
Iant\-chen\-ko and
Korotyaev \cite{IK14} considered the Dirac operator on the real line
with 2x2 matrix-valued compactly supported potentials. They obtained
asymptotics of counting function of resonances, estimates on the
resonances and the forbidden domain, a trace formula in terms of
resonances. Lieb-Thirring type inequality for resonances  of Dirac
operators with compactly supported matrix-valued potentials on the
real line is obtained in \cite{K14}. Resonances  for
 Stark operators on the real line are considered in \cite{K16x}.
Here we underline that for all these cases the
 corresponding Riemann surfaces are two-sheeted similar to the
 Schr\"odinger operator case.

A lot of papers are devoted to the inverse scattering theory for
fourth order operators on the line, see papers Aktosun and
Papanicolaou \cite{AP08}, Butler \cite{Bu68}, Iwasaki \cite{I88},
\cite{Iw88}, Hoppe, Laptev and \"Ostensson \cite{HLO06} and the book
Beals, Deift, Tomei \cite{BDT88} and references therein.

Resonances for higher order operators with compactly supported
coefficients were considered by Korotyaev \cite{K16} firstly for the
case of third order operators. Here general properties of resonances
were described. In particular, upper bounds of the number of
resonances in complex discs at large radius and the trace formula in
terms of resonances were obtained. Note that this case is very
complicated for the consideration since the Born term roughly
speaking is constant. Recall that for Schr\"o\-dinger operators the
corresponding Born term is the Fourier transformation of the
potential. It is important for the global analysis of resonances,
including inverse problems.


Resonances of fourth order operators with compactly
supported coefficients on the half-line were studied by Badanin and
Korotyaev \cite{BK17}. Asymptotics of resonances and trace formulas
in terms of resonances were determined. This case is simpler, than
the case of the line considered in the present paper,
because the scattering matrix is a scalar
function. An
extension of the determinant onto the third quadrant $\K_3$ may be
obtained using a $2\ts 2$ matrix-valued function $\O$.
 Here the technique from \cite{K16} was used.
 But it is
important that the corresponding Born term is expressed in terms of
the Fourier
transformations of the compactly supported coefficients.

In the present paper the
corresponding matrix $\O$ for an operator on the line is a $4\ts 4$
matrix-valued function and there are some algebraic difficulties in
order to obtain this extension. Clearly, the problem for higher
order operators will be much more complicated, especially for odd
order case.

The usual applications of fourth order differential operators
are bending vibrations of thin beams and plates described
by the Euler-Bernoulli equation.
Many problems of engineering involve
solutions of scattering problem for the Euler-Bernoulli equation,
see \cite{Gr75} and references therein.
Furthermore, the inverse spectral problem methods for some
non-linear partial differential equations
lead to fourth order operators, see \cite{HLO06}.

\medskip

The plan of the paper is as follows.
In Section~2 we study properties
of the resolvent of the operator $H_0$.
In Sections~3 and 4 we consider the scattering matrix and the determinant.
In Sections~5 we prove Proposition~\ref{T1}
and Theorems~\ref{CorEstN}, \ref{ThRes}.
Moreover, there
we consider the Euler-Bernoulli operator
and prove Corollary~\ref{CorEB} and Theorem~\ref{ThEB}.
In Section~6 we obtain asymptotics of the resonances and
prove Theorem~\ref{ThAsCF}.

\section {Properties of the free resolvent}
\setcounter{equation}{0}

\subsection{The well-known facts}
By $\cB$ we denote the class of bounded operators.
Let $\cB_1$ and $\cB_2$ be the trace and
the Hilbert-Schmidt class equipped with the norm $\|\cdot \|_{\cB_1}$
and $ \|\cdot \|_{\cB_2}$ correspondingly.
We recall some well known facts. Let $A, B\in \cB$ and $AB, BA,X\in\cB_1$. Then
\[
\lb{2.1} \Tr AB=\Tr BA,
\]
\[
\lb{2.2} \det (I+ AB)=\det (I+BA),
\]
\[
\lb{B1}
|\det (I+ X)|\le e^{\|X\|_{\cB_1}},
\]
see e.g., Sect.~3. in the book \cite{S05}.
Let  the operator-valued function $X :\cD\to \cB_1$ be analytic for
some domain $\cD\ss\C$ and $(I+X (z))^{-1}\in \cB$ for any $z\in
\cD$. Then for the function $F(z)=\det (I+X (z))$ we have
\[
\lb{2.3}
F'(z)= F(z)\Tr (I+X (z))^{-1}X '(z),\qqq z\in \cD.
\]

Introduce the space $L^p(\R)$ equipped by the norm
$\|f\|_p=(\int_\R|f(x)|^p dx)^{1\/p}\ge 0$ and
we write $\|f\|=\|f\|_2$.

\subsection{Schr\"odinger operator}
We discuss a Schr\"odinger operator $h$ on $L^2(\R)$ given by
$$
h=h_0-p,\qqq h_0=-{d^2\/dx^2},
$$
 where $h_0$ is the unperturbed operator and the potential
 $p\in\cH_0$.

$\bu$
The operator $r_0(k)=(h_0-k^2)^{-1}, k\in \C_+$,
is an integral operator having
the kernel $r_0(x-x',k), x,x'\in \R$ given by
\[
\lb{h1}
r_0(x,k)={1\/2\pi}\int_\R {e^{itx}\/t^2-k^2}dt
={ie^{ik|x|}\/2k}.
\]
Define an operator-valued function $g(k)=\vt r_0(k)\vp, k\in \C_+$, where
$\vt^2, \vp^2\in \cH_0$. For each $k\in \C_+$ the operator $g(k)\in \cB_j, j=1,2$
and the mappings
\[
\lb{ga}
g:\C_+\to \cB_j,
\]
is analytic and it has an analytic extension into the whole complex plane without zero.
Thus the operator-valued
function $g: \C\sm \{0\}\to \cB_j$ is analytic.
Moreover, we have the following estimate
\[
\lb{arbB2}
\begin{aligned}
\|g(k)\|_{\cB_2}\le {\|\vt\|\|\vp\|\/2|k|},\qqq k\in \ol\C_+\sm\{0\}.
\end{aligned}
\]
Define the Fourier transformation $\F: L^2(\R)\to L^2(\R)$ by
\[
\lb{Ft}
\begin{aligned}
\F f(\x)={1\/\sqrt{2\pi}}\int_\R f(x)e^{-i\x x}dx,\ \ \
\x\in\R.
\end{aligned}
\]
Then $r_0(k)=\F^* \e_{-k}\e_k\F$, where $\e_k(\x)$
is the multiplication by
$(\x-k)^{-1}$ and we have
\[
\lb{arb1}
\begin{aligned}
\|g(k)\|_{\cB_1}=\big\|\vt (h_0-k^2)^{-1}\vp\big\|_{\cB_1}
\le\big\|\vt \F^* \e_{-k}\big\|_{\cB_2}
\big\|\e_{k}\F\vp \big\|_{\cB_2}
\\
\le
{\|\vt\|\|\vp\|\/2\pi}\|\e_{-k}\|\|\e_{k}\|=
{\|\vt\|\|\vp\|\/2\Im k},\qqq \Im k>0,
\end{aligned}
\]
since $\int_\R|\x\pm k|^{-2}d\x={\pi\/\Im k}$.

$\bu$  The Schr\"odinger equation
$-y''-py=k^2 y,k\in\C\sm\{0\}$, has unique Jost solutions
$f_{\pm}(x,k)$ satisfying the conditions
$f_+(x,k)=e^{ikx},x>\g$ and $f_-(x,k)=e^{-ikx},x<0$.
For each $x\in \R$  the function $f_\pm(x,k)$ is entire.
The following identity holds true:
$$
f_+(x,k)=b(k)f_-(x,k)+a(k)f_-(x,-k),\qq k\in\C\sm\{0\},
$$
where the functions $a,b$ are defined by
$$
 a(k)={\{f_-(x,k),f_+(x,k)\}\/2ik},\qqq
b(k)={\{f_+(x,k),f_-(x,-k)\}\/2ik},
$$
and $\{f,g\}=fg'-f'g$ denotes the Wronskian.
The scattering matrix for the pair $h,h_0$
has the following form
$$
s(k)=\ma a(k)^{-1}& r_-(k)\\r_+(k)&a(k)^{-1}\am,\qq
r_{\pm}(k)=\pm {b(\mp k)\/a(k)},
$$
where $a^{-1}$ is the transmission coefficient and
$r_\pm$ are the reflection coefficients.
It is well known the following identity
$a(k)=d(k)$ for all $ k\in \C$, where $d(k)$
is the determinant defined by \er{defd}.
Moreover, the scattering matrix satisfies
\[
\lb{SD2od}
\det s(k)={d(-k)\/d(k)}\qqq\forall\qq k>0.
\]

$\bu$
The function $d(k)$ satisfies
$$
d(k)=1+O(k^{-1})\qq\as\qq|k|\to\iy,\qq k\in\C_+,
$$
uniformly on $\arg k\in[0,\pi]$. Then we can define the function
$\log d(k)$ by the condition $\log d(it)=o(1)$ as $t\to\iy$, which
satisfies
$$
i\log d(k)={1\/2k}\Big(\int_\R p(x)dx+o(1)\Big)
\qq \as \qq\Im k\to+\iy.
$$

\subsection{The free resolvent}
We rewrite the free resolvent $R_0(k)=(H_0-k^4)^{-1}$ in terms of
the resolvent $r_0(k)=(h_0-k^2)^{-1}$ by
\[
\lb{R0r0}
R_0(k)={r_0(k)-r_0(ik)\/2k^2},\qqq
\pa R_0(k)\pa=-{r_0(k)+r_0(ik)\/2},\qqq \forall \ k\in \K_1.
\]
Then the kernel of the free resolvent
$R_0(k)$ has the form $R_0(x-x',k),  x,x'\in\R$, where
\[
\lb{R01}
R_0(x,k)={ie^{ik|x|}-e^{-k|x|}\/4k^3},
\]
and satisfies
\[
\lb{asR010}
R_0(x,k)={i-1\/4k^3}-{(1+i)x^2\/4k}+O(1)\qqq\as\qq |k|\to 0,
\]
locally uniformly in $x\in \R$.
Each function $R_0(x,k)-{i-1\/4k^3}+{(1+i)x^2\/4k},x\in\R$,
is entire in $k$.

Define the operator-valued function
$G(k)=\vt \pa R_0(k)\vp, k\in \K_1,$
where $\vt^2,\vp^2\in\cH_0$.
The identity \er{R0r0} yields that
for each $k\in \K_1$ the operator
$G(k)\in\cB_j,j=1,2$ and the mappings
\[
\lb{Ga}
G(k): \K_1\to \cB_j
\]
are analytic and they have analytic extensions into whole complex
plane without zero.
Moreover, from \er{R01} we have the following estimate
$$
\|G(k)\|_{\cB_2}\le{\|\vt\|\|\vp\|\/|k|^2},\qqq k\in \ol\K_1\sm\{0\}.
$$
Moreover, we obtain
$\pa R_0(k)=\F^* \r_{ik}\r_ks\F$, where $\r_k(\x)$
is the multiplication by
$(\x^2-k^2)^{-1}|\x|^{1\/2}$, $s$ is the multiplication by $\sign\x$
 and we have
\[
\lb{arb2}
\|G(k)\|_{\cB_1}\le
\big\|\vt\F^* \r_{ik}\big\|_{\cB_2}\big\|\r_{k}\F\vp\big\|_{\cB_2}
\le
{\|\vt\|\|\vp\|\/2\pi}\|\r_{ik}\|\|\r_{k}\|=
{\|\vt\|\|\vp\|\/4\Re k\Im k},
\]
$k\in\K_1,$
since $\int_\R |\x||\x^2\pm k^2|^{-2}d\x={\pi\/4\Re k\Im k}$.

\subsection{Resolvent estimates}
The operator $H_0$ is self-adjoint  on the form domain given by
$
\mD(H_0)=\{y,y''\in L^2(\R)\}.
$
The quadratic form $(Vy,y)$ is defined by
$(Vy,y)=-(py',y')+(qy,y)$, $y\in\mD(H_0)$, where
$(u,v)$ is the scalar product in $L^2(\R)$.
Then the standard arguments (see e.g., \cite{K03}) give
\[
\lb{VE}
|(Vy,y)|\le{1\/2}\|y''\|^2+C\|y\|^2\qqq
\forall \qq y\in\mD(H_0)
\]
for some constant $C>0$.
Then the KLMN Theorem (see \cite[Th~X.17]{RS75}) yields
that there exists a unique self-adjoint operator  $H=H_0+V$
with the form domain $\mD(H)=\mD(H_0)$ and
\[
\lb{qf4}
(Hy,y_1)=(H_0y,y_1)+(Vy,y_1)\qq\forall\qq
y,y_1\in\mD(H_0).
\]

In order to study the determinant we need to consider $Y_0$.
The definitions \er{defVj}, \er{Y0} imply
\[
\lb{exrepY0}
Y_0 = \ma (2p)^{1\/2}\pa R_0\pa |2p|^{1\/2}&
(2p)^{1\/2}\pa R_0|q|^{1\/2}\\
q^{1\/2} R_0\pa |2p|^{1\/2}&q^{1\/2} R_0 |q|^{1\/2}\am.
\]
We introduce the operator-valued function $Y$ by
\[
\lb{defJJ0}
Y(k) =V_2R(k)V_1,\qqq k\in \K_1.
\]
This function satisfies the standard identity
\[
\lb{2.5}
(I-Y(k))(I+Y_0(k))=I\qqq \forall \ k\in \K_1\sm \s_d,
\]
where $\s_d$ is the set of the zeros
of the function $D$ in $\ol\K_1$.

\begin{lemma}
\label{TY}
Let $p,q\in\cH_0$. Then

i) The operator $Y_0(k)\in \cB_j, j=1,2$ for each $k\in \K_1$,
the operator-valued
function $Y_0: \K_1\to \cB_j$ is analytic and has an analytic
extension into the whole complex plane without zero.
The
operator-valued function $k^3Y_0(k)$ is entire.
Moreover, $Y_0$ satisfies
\[
\lb{estY0op}
\|Y_0(k)\|_{\cB_2}\le{C\/|k|},
\]
\[
\lb{estY0}
\begin{aligned}
\|Y_0(k)\|_{\cB_1}\le
(\|2p\|+\|q\|)\Big({1\/\Re k}+{1\/\Im k}\Big)\Big(1+{1\/|k|}\Big)^2,
\end{aligned}
\]
$k\in\K_1,$ for some constant $C=C(p,q)>0$.

ii) The operator $Y(k)\in\cB_1$ for each
$k\in\K_1\sm \s_d$ and  the operator-valued
function $Y: \K_1\sm \s_d\to \cB_1$ is analytic and has a meromorphic
extension from $\K_1\sm \s_d$ into the whole complex plane.
Moreover, $Y$ satisfies
\[
\lb{Y2}
\|Y(k)\|_{\cB_2}={O(1)\/|k|},
\]
\[
\lb{estY2}
\|Y(k)-Y_0(k)\|_{\cB_2}= {O(1)\/|k|^2},
\]
as $ k\in\ol\K_1, |k|\to\iy$ and  uniformly in $\arg k\in[0,\pi]$.
\end{lemma}

{\bf Proof.}
i) Substituting the identity \er{R01} into the definition \er{Y0}
we obtain \er{estY0op}.
Substituting the identities \er{R0r0}
into \er{exrepY0} and using the facts about the mappings $g,G$
in \er{ga}, \er{Ga}  we deduce that
 the operator-valued
function $Y_0: \K_1\to \cB_1$ is analytic and has an analytic
 extension into the whole complex plane without zero.
 The asymptotics \er{asR010} shows that the
operator-valued function $k^3Y_0(k)$ is entire.

Using  the estimates \er{arb1} we obtain for $\Im k>0$:
$$
\begin{aligned}
\|p^{1\/2}r_0(k)|p|^{1\/2}\|_{\cB_1}\le {\|p\|_1\/2\Im k},\qqq \qqq
\|q^{1\/2}r_0(k)|q|^{1\/2}\|_{\cB_1}\le
{\|q\|_1\/2\Im k},\\
\|p^{1\/2}r_0(k)|q|^{1\/2}\|_{\cB_1}\le
{(\|p\|_1\|q\|_1)^{1\/2}\/2\Im k},
\end{aligned}
$$
 and the similar estimates with $r_0(ik)$ .
These estimates and the relations \er{arb2}, \er{exrepY0} give
$$
\|Y_0(k)\|_{\cB_1}\le {1\/4}(\|2p\|_1^{1\/2}+\|q\|_1^{1\/2})^2
\Big({1\/\Re k}+{1\/\Im k}\Big)\Big(1+{1\/|k|}\Big)^2,
$$
which yields \er{estY0}.

ii) For $k\in\K_1\sm \s_d$ identity \er{2.5} gives
\[
\lb{idYY0}
Y(k)=I-(I+Y_0(k))^{-1}=Y_0(k)(I+Y_0(k))^{-1}\in\cB_j,\qq j=1,2,
\]
and, since $Y_0(k)$ is analytic in $\K_1$, $Y(k)$ is analytic
in $\K_1\sm \s_d$.
Due to the analytic Fredholm theorem, see \cite[Th VI.14]{RS72},
the function $Y(k)$ has a meromorphic  extension into the
whole complex plane.
The estimate \er{estY0op} implies the asymptotics \er{Y2}.
Moreover,
$$
Y(k)-Y_0(k)=Y_0(k)\big((I+Y_0(k))^{-1}-I\big),
$$
which yields \er{estY2}.
\BBox

\section{The scattering matrix.}
\setcounter{equation}{0}

\subsection{The spectral representation for $H_0$}
Define a unitary operator
$$
\cF:L^2(\R)\to L^2(\R_+,dk)\os L^2(\R_+,dk)
$$
by
\begin{equation}
\label{defG}
\cF f(k)=\ma\wh f(k)\\ \wh f(-k)\am,\qq k\in\R_+,\qq
\wh f(k)=\F f(k)={1\/\sqrt{2\pi}}\int_\R f(x)e^{-ik x}dx.
\end{equation}
The identity $(\F H_0\F^*\wh f)(k)=k^4\wh f(k),\ k\in \R$,
implies that $\cF H_0\cF^*$ is the operator of multiplication by $k^4$
in $L^2(\R_+,dk)\os L^2(\R_+,dk)$.

Introduce the operators $\p_1(k):L^2(\R)\os L^2(\R)\to \C^2$
and  $\p_2(k):\C^2\to L^2(\R)\os L^2(\R)$ for each $k>0$ by
\[
\lb{defPsi}
\begin{aligned}
\p_1(k)f=\ma\p_{11}(k)\\\p_{21}(k)\am f=\cF V_1 f(k)=\ma\F V_1f(k)\\
\F V_1 f(-k)\am =\int_\R\p_1(x,k)f(x)dx,
\\
(\p_2(k)c)(x)=\big(\p_{12}(k),\p_{22}(k)\big)c
={1\/\sqrt{2\pi}}V_2(e^{ikx},e^{-ikx})c=\p_2(x,k)c,
\end{aligned}
\]
where $f\in L^2(\R)\os L^2(\R)$ and $ c\in \C^2$. Here
$\p_{j1}(k):L^2(\R)\os L^2(\R)\to \C$ and $\p_{j2}(k):\C\to
L^2(\R)\os L^2(\R), j=1,2$, and the kernels
$\p_j(\cdot,\cdot),j=1,2$ have the forms:
\[
\lb{dpxk}
\begin{aligned}
\p_1(x,k)={1\/\sqrt{2\pi}}
\ma ik|2p(x)|^{1\/2}e^{-ikx}&|q(x)|^{1\/2}e^{-ikx}\\
-ik|2p(x)|^{1\/2}e^{ikx}&|q(x)|^{1\/2}e^{ikx}\am,
\\
\p_2(x,k)={1\/\sqrt{2\pi}}
\ma ik(2p(x))^{1\/2}e^{ikx}&-ik(2p(x))^{1\/2}e^{-ikx}\\
q^{1\/2}(x)e^{ikx}&q^{1\/2}(x)e^{-ikx}\am.
\end{aligned}
\]
It is clear that
the operator-valued functions $\p_j(k), k\in\R_+, j=1,2$
have analytic extensions from $\R_+$ into the whole complex plane.
Then we can introduce the operators $\P_1(k)$
and $\P_2(k)$ by
\[
\lb{defPsi12}
 \P_1(k)=\ma\p_1(ik)\\ \p_1(k)\am,\qqq
 \P_2(k)=\Big(i\p_2(ik), \p_2(k)\Big),\qqq k\in\C.
\]

\begin{lemma}
\label{TPk}
Let $p,q\in\cH_0$.
Then  the operator-valued functions $\p_j(k),j=1,2,$
are entire and
satisfy
\[
\label{Pe}
\|\p_{jj}(k)\|\le C(k)e^{\g (\Im k)_+},\qq \|\p_{j\ell}(k)\|\le C(k)e^{\g (\Im k)_-}
\qq
\]
for all $k\in\C, j,\ell=1,2, j\ne\ell$, where
$$
(a_\pm)=\max\{\pm a,0\},\qq a\in\R,\qqq C(k)={1\/\sqrt{2\pi}}
\big(2|k|^2\|p\|_1+\|q\|_1\big)^{1\/2}.
$$
\end{lemma}

 {\bf Proof.}  The operator $\p_1(k), k\in \R_+$ has
the kernel $\p_1(x,k)$ given by \er{dpxk}.
Let $g=\p_1(k)f,f=(f_1,f_2)\in L^2(\R)\os L^2(\R)$.
We have
$$
g=\ma g_+\\ g_-\am =\int_0^\g\p_1(x,k)f(x)dx={1\/\sqrt{2\pi}}
\int_0^\g\ma e^{-ikx} \big(ik |2p(x)|^{1\/2}f_1(x)+|q(x)|^{1\/2}f_2(x)\big)\\
e^{ikx} \big(-ik|2p(x)|^{1\/2}f_1(x)+|q(x)|^{1\/2}f_2(x))\am dx,
$$
and
$$
|g_\pm|^2\le e^{2\g (\Im k)_\pm}{1\/2\pi}\|f\|^2
\int_0^\g \big(|k|^2|2p(x)|+|q(x)|\big)dx.
$$
This yields the estimates \er{Pe} for $\p_1(k)$.
The proof for $\p_2(k)$ is similar.
\BBox

\medskip

Introduce finite rank operators $P_1(k),P_2(k),k\in\C\sm\{0\}$,
acting on $L^2(\R)\os L^2(\R)$,
by
\[
\lb{defP}
\begin{aligned}
P_1(k)=c_k\p_2(k)\p_1(k), \qqq c_k={\pi\/i2k^3},
\\
P_2(k)=P_1(ik)+P_1(k)=c_k \P_2(k)\P_1(k).
\end{aligned}
\]
The operators
$P_1(k),P_2(k)$ are analytic
in the domain $\C\sm \{0\}$.
Below we need the following simple identities.

\begin{lemma}
Let $p,q\in\cH_0$. Then for any $k\in\C\sm\{0\}$
the operators
$P_1(k),P_2(k)$  satisfy
\[
\lb{cP1}
P_1(k)=Y_0(ik)-Y_0(k),
\]
\[
\lb{cP2}
P_2(k)= Y_0(-k)-Y_0(k).
\]
\end{lemma}

\no {\bf Proof.}
The  identities \er{h1}, \er{R0r0}
yield
\[
\lb{Y0-Y0}
R_0(y,k)-R_0(y,ik)={r_0(y,k)-r_0(y,-k)\/2k^2}={i\cos ky\/2k^3},
\qqq \forall \qq (k,y)\in \C\ts \R,\qq k\ne 0.
\]
For all $f\in L^2(\R)\oplus L^2(\R)$
the definitions \er{defPsi} imply
$$
\p_2(k)\p_1(k)f={1\/2\pi}V_2(e^{ikx},e^{-ikx})
\int_\R\ma e^{-ikx'}\\ e^{ikx'}\am V_1f(x')dx'
={1\/\pi}V_2\int_\R\cos k(x-x')V_1f(x')dx'.
$$
This identity together with
the identity \er{Y0-Y0} and the definition \er{Y0}
gives
$$
Y_0(ik)-Y_0(k)=V_2(R_0(ik)-R_0(k))V_1
=c_k\p_2(k)\p_1(k),
$$
which yields the identity \er{cP1}.

The identities
$$
Y_0(-k)-Y_0(k)=Y_0(-k)-Y_0(ik)+Y_0(ik)-Y_0(k)=P_1(ik)+P_1(k)
$$
give \er{cP2}.
\BBox

\subsection{The scattering matrix.}
We define the S-matrix for the operators $H_0,H$.
It is well known that the wave operators $W_\pm=W_\pm(H,H_0)$ for
the pair $H_0, H$, given by
$$
W_\pm=s-\lim e^{itH}e^{-itH_0} \qqq \as \qqq t\to \pm\iy,
$$
exist and  are complete, i.e., $\Ran W_\pm=\mH_{ac}(H)$.
The scattering operator $\cS=W_+^*W_-$ is unitary. The operators $H_0$
and $\cS$ commute and thus are simultaneously diagonalizable:
\[
\lb{DLH0}
L^2(\R)=\int_{\R_+}^\oplus \mH_\l d\l,\qqq
H_0=\int_{\R_+}^\oplus\l I_\l d\l,\qqq \cS=\int_{\R_+}^\oplus
S(\l^{1\/4})d\l;
\]
here $I_\l$ is the identity in the fiber space $\mH_\l=\C^2$ and
$S(k),k=\l^{1\/4}>0$
is the scattering matrix
(which is a  $2\ts 2$ matrix-valued function of $k>0$ in our case)
for the pair $H_0, H$.

The operator $\cS =W_+^*W_-$ commutes with the operator
$H_0$ and the operator $\cF H_0\cF^*$ is the operator of
multiplication by $k^4$ in $L^2(\R_+,dk)\os L^2(\R_+,dk)$.
Then the operator
$\cF\cS\cF^*$ acts in the space $L^2(\R_+,dk)\os L^2(\R_+,dk)$
as multiplication
by a $2\ts 2$ matrix-valued function $S(k)$.

The scattering matrix $S(k)$ is a continuous $2\ts 2$ matrix-valued
function in $\R_+\sm\s_d$, where $\s_d$ is the set of the zeros
of the function $D$ in $\ol\K_1$, and satisfies
\[
\label{S}
S(k)=\ma s_{11}(k)&s_{12}(k)\\
s_{21}(k)&s_{22}(k)\am=\1_2+c_k\cA(k),
\qqq\forall\ k=\l^{1\/4}>0, \qqq c_k={\pi\/i2k^3}
\]
(see, e.g., \cite{RS79}),
where  $\1_N$ is the $N\ts N$ identity matrix and    $\cA(k)$ is
the scattering amplitude given by
\[
\lb{ScA}
\cA(k)
=\cA_0(k)-\cA_1(k),
\]
where the Born term $\cA_0(k)$ and the term $\cA_1(k)$
have the form
\[
\lb{ScA01}
\cA_0(k)=\p_1(k)\p_2(k),\qqq
\cA_1(k)=\p_1(k)Y(k)\p_2(k).
\]
The operator-valued functions $\p_1(k),\p_2(k)$ are analytic
in $\C$, then
the matrix-valued function $\cA_0(k)$ has
an analytic extension from $\R_+$
into the whole complex plane.

\subsection{The scattering amplitude}
Now we consider the scattering amplitude $\cA$.

\begin{lemma}
\label{TA}
i) Let $p,q\in\cH_0$.
Then the scattering amplitude $\cA(k), k\in \R_+$
has a meromorphic extension from $\R_+$
into the whole complex plane.
Moreover, the matrix-valued function $\cA_0(k)$
is entire and satisfies
\[
\label{A0i}
\cA_0(k)={1\/2\pi} \ma
\a_1(k)&\a_2(k)\\
\a_2(-k)&\a_1(k)\am ,
\]
for all $k\in\K_1$,
where
\[
\lb{al12}
\a_1(k)=q_0-2k^2p_0,\qqq
\a_2(k)=\sqrt{2\pi}\big(2k^2\wh p(2k)+\wh q(2k)\big),
\]
$$
p_0=\int_\R p(x)dx,\qq q_0=\int_\R q(x)dx,\qq
\wh f(k)={1\/\sqrt{2\pi}}\int_\R f(x)e^{-ikx}dx,
$$
the matrix-valued functions $\cA_0(k),\cA_1(k)$
satisfy
\[
\lb{asA0r}
\cA_0(k)=\ma O(k^2)&e^{2\g\Im k}o(k^2)\\o(k^2)&O(k^2)\am,\qq
\cA_0(ik)=\ma O(k^2)&e^{2\g\Re k}o(k^2)\\o(k^2)&O(k^2)\am,
\]
\[
\label{asA1xim}
\cA_1(k)=\ma e^{\g\Im k}O(k)&e^{2\g\Im k}O(k)\\
O(k)&e^{\g\Im k}O(k)\am,
\]
as $|k|\to\iy,k\in\ol\K_1$ uniformly in $\arg k\in [0,{\pi\/2}]$.

ii) Let $(p,q)\in\cH_1\ts\cH_0$.
Then the functions $\cA_0(k),\cA_1(k)$ satisfy
\[
\lb{asA0}
\cA_0(k)={k\/2\pi} \ma
-2k(p_0+O(k^{-2}))
&ie^{-i2k\g}(p_++o(1)+e^{-2\g\Im k}O(1))\\
ip_-+o(1)+e^{-2\g\Im k}O(1)&
-2k(p_0+O(k^{-2}))\am,
\]
\[
\lb{ascA0ik}
\cA_0(ik)=-{k\/2\pi} \ma
2k(p_0+O(k^{-2}))
&e^{2k\g}(p_++o(1)+e^{-2\g\Re k}O(1))\\
p_-+o(1)+e^{-2\g\Re k}O(1)&
2k(p_0+O(k^{-2}))\am,
\]
\[
\lb{ascA1im}
\cA_1(k)=\ma e^{\g\Im k}O(1)&e^{2\g\Im k}O(1)\\
O(1)&e^{\g\Im k}O(1)\am,
\]
as $|k|\to\iy, k\in\ol\K_1$, uniformly in $\arg k\in[0,{\pi\/2}]$,
where
$
p_+=p(\g-0), p_-=p(+0).
$
\end{lemma}

\no {\bf Proof.}
i) The operator-valued functions $\p_1,\p_2$ are entire, then
the function $\cA_0(k)$ has an analytic
extension from $\R_+$ into the whole complex plane.
Due to Lemma \ref{TY}~ii), the function
$\cA_1(k)$ has a meromorphic extension from $\R_+$
onto the whole complex plane. Then
the scattering amplitude
$\cA(k)$ has
a meromorphic extension from $\R_+$ into the whole complex plane.

The definitions \er{defPsi} and \er{ScA01} give
\[
\lb{idcA0}
\cA_0(k)=\int_\R\p_1(x,k)\p_2(x,k) dx,
\]
\[
\lb{idcA1}
\cA_1(k)=\ma\p_{11}(k)\\\p_{21}(k)\am Y(k)
\Big(\p_{12}(k),\ \p_{22}(k)\Big).
\]
Substituting the identities \er{dpxk} into \er{idcA0}
we obtain the identity \er{A0i}, which yields the asymptotics
\er{asA0r}.
The estimates \er{Y2} and \er{Pe}
and the identity \er{idcA1} give the asymptotics \er{asA1xim}.

ii) Let $|k|\to\iy, k\in\ol\K_1$. The integration by parts gives
$$
\begin{aligned}
\a_2(k)=\int_0^\g(2k^2p(x)+q(x))e^{-2ikx}dx
=ik(p_++o(1))e^{-i2k\g}+O(k),
\\
\a_2(-k)=\int_0^\g(2k^2p(x)+q(x))e^{2ikx}dx
=ik(p_-+o(1))+e^{i2k\g}O(k),
\end{aligned}
$$
Substituting these asymptotics and the definition \er{al12} into
the identity \er{A0i} we obtain \er{asA0}.
Similarly,
$$
\begin{aligned}
\a_2(ik)=\int_0^\g(-2k^2p(x)+q(x))e^{2kx}dx
=-k(p_++o(1))e^{2k\g}+O(k),
\\
\a_2(-ik)=\int_0^\g(-2k^2p(x)+q(x))e^{-2kx}dx
=-k(p_-+o(1))+e^{-2k\g}O(k),
\end{aligned}
$$
which yields \er{ascA0ik}.

The definition \er{ScA01} and the estimates \er{estY2} and \er{Pe}
give
\[
\lb{asA1pr}
\cA_1(k)=\p_1(k)Y_0(k)\p_2(k)
+\ma e^{\g\Im k}O(1)&e^{2\g\Im k}O(1)\\
O(1)&e^{\g\Im k}O(1)\am.
\]
The identities \er{exrepY0} and \er{dpxk} imply
$$
\begin{aligned}
\p_{11}(k)Y_0(k)\p_{12}(k)={1\/2\pi}\iint_{[0,\g]^2}
e^{-ik(x-x')}\Big(-4k^2p(x)p(x'){\pa^2 R_0(x,x',k)\/\pa x^2}
\\
+2ik\big(p(x)q(x')
+p(x')q(x)\big){\pa R_0(x,x',k)\/\pa x}
+q(x)q(x')R_0(x,x',k)\Big)
dxdx'.
\end{aligned}
$$
Substituting the kernel \er{R01} into the last identity
and integrating by parts in the first term we obtain
$$
\p_{11}(k)Y_0(k)\p_{12}(k)=e^{\g\Im k}O(1).
$$
Similarly,
$$
\begin{aligned}
\p_{11}(k)Y_0(k)\p_{22}(k)&=e^{2\g\Im k}O(1),\qqq
\p_{21}(k)Y_0(k)\p_{12}(k)=O(1),\\
&\p_{21}(k)Y_0(k)\p_{22}(k)=e^{\g\Im k}O(1),
\end{aligned}
$$
which yields
$$
\p_1(k)Y_0(k)\p_2(k)
=\ma e^{\g\Im k}O(1)&e^{2\g\Im k}O(1)\\
O(1)&e^{\g\Im k}O(1)\am.
$$
Substituting this asymptotics into \er{asA1pr} we obtain
the asymptotics \er{ascA1im}.
\BBox

\section{The determinant}
\setcounter{equation}{0}

\subsection{Asymptotics of the determinant}
Lemma~\ref{TY}~i) shows that
$Y_0(k)\in\cB_1$, then the determinant
$D(k)=\det(I+Y_0(k)), k\in\K_1$ is well defined.

\begin{lemma}
\lb{TD1}
Let $p,q\in\cH_0$. Then

i) The determinant $D(k)=\det
(I+Y_0(k))$ is analytic in  $\K_1$ and  has  an analytic  extension
from $\K_1$ into the whole complex plane without zero,
such that the function $k^4D(k)$ is entire.

ii)  The function $D(k)$ is real on the line $e^{i{\pi\/4}}\R$.
\end{lemma}

\no {\bf Proof.} i) Due to Lemma~\ref{TY}~i) the operator-valued
function $Y_0(k)$, and then the determinant $D(k)$, is analytic in
$k\in \K_1$ and has  an analytic extension from $k\in \K_1$ into the
whole complex plane without zero.
It is proved in \cite{BK16} that the function $k^4D(k)$ is entire,
the proof is rather technical.

ii) The identity \er{R01} shows that $R_0(k)$
is real on the line
$e^{i{\pi\/4}}\R$, then $Y_0(k)$ is real also.
Therefore, $D(k)$ is real on this line.
\BBox

\medskip

The identities \er{h1}, \er{R0r0}, \er{exrepY0} imply
\[
\lb{TrY0}
\begin{aligned}
\Tr Y_0(k)
=\int_\R\Big(-\big(r_0(x,x,k)+r_0(x,x,ik)\big)p(x)
+{(r_0(x,x,k)-r_0(x,x,ik))q(x)\/2k^2}\Big)dx
\\
=-{(1+i)p_0\/2k}-{(1-i)q_0\/4k^3},
\end{aligned}
\]
where $p_0=\int_\R p(x)dx,q_0=\int_\R q(x)dx.$
The estimates \er{estY0} give $\|Y_0(k)\|_{\cB_1}=O(k^{-1})$ as
$k\to e^{i{\pi\/4}}\infty$.
We can define the branch $\log D(k)$, for $k\in\K_1$
and $|k|$ large enough, by
$$
\log D(k) = o(1)\qqq
\as\qqq k\to e^{i{\pi\/4}}\infty.
$$
We need the following standard results.

\begin{lemma}
\lb{TD2}
Let $p,q\in\cH_0$.
Then
the function $\log D(k)$ satisfies
\[
\lb{2.13}
 |\log D(k)+\sum _{n=1}^{N}{1\/n}\Tr (-Y_0(k))^n|\le
{C_1\/|k|^{N+1} },\qqq \forall  \qq N\ge 1,
\]
\[
\lb{2.12}
\log D(k)=-\sum _{n=1}^\iy {1\/n}\Tr (-Y_0(k))^n, \ \ \
\]
for any $k\in \K_1, |k|$ large enough, and for some $C_1>0$,
where the series converges absolutely and uniformly in $k$.
Furthermore, the function $\log D$ satisfies the asymptotics
 \[
\lb{aD1}
 \log D(k)=-\fr{(1+i)p_0}{2k}+{O(1)\/k^2}
\qqq \as \qq |k|\to \iy,\qq k\in \ol\K_1
\]
uniformly in $\arg k\in[0,{\pi\/2}]$.
\end{lemma}

\no {\bf Proof.} Let $k\in\K_1$. The estimate \er{estY0op} gives
\[
\lb{estY0n}
\big|\Tr(Y_0(k))^n\big|\le\|Y_0(k)\|_{\cB_2}^{n}
\le\Big({C\/|k|}\Big)^{n},\qqq n\ge 2.
\]
Then the series \er{2.12} converges absolutely
and uniformly and
it is well-known that
the sum is equal to $\log D(k)$
(see \cite[Lm XIII.17.6]{RS78}).
Using the estimates \er{estY0n} we obtain \er{2.13}.
The estimate \er{2.13} together with
the identity \er{TrY0} gives the asymptotics \er{aD1}.
\BBox

\subsection{Identities for the determinant and S-matrix}
Asymptotics of the determinant in $\C_-$
in the case of the Schr\"odinger
operator is obtained from the asymptotics in $\C_+$
and the identity \er{SD2od}.
In order to determine asymptotics of the determinant
in $\K_2,\K_3,\K_4$ for the case of fourth order operators
we need some additional
identities.
The situation for third
order operators is described in \cite{K16}.

Recall that the $S$-matrix $S(k)$ is a meromorphic
matrix-valued function and satisfies the identity
$
S(k)=\1_2+c_k\big(\cA_0(k)-\cA_1(k)\big),
$
see \er{S}, \er{ScA},
where
$
\cA_0(k)=\p_1(k)\p_2(k)
$
is the Born approximation for the scattering
amplitude $\cA=\cA_0-\cA_1$,
$
\cA_1(k)=\p_1(k)Y(k)\p_2(k).
$

Introduce the $4\ts 4$ matrix-valued function
$\O(k),k\in\K_1\sm\s_d$ by
\[
\lb{defOmega}
\O(k)=\1_4+c_k\big(\O_0(k)-\O_1(k)\big), \qqq c_k={\pi\/i2k^3},
\]
where the ``Born'' term $\O_0(k)$ has the form
\[
\lb{defOm0}
\O_0(k)=\P_1(k)\P_2(k),
\]
$\P_1,\P_2$ are given by \er{defPsi12}, and
\[
\lb{defOm11}
\O_1(k)=\P_1(k)Y(k)\P_2(k),
\]
The function $\O_0(k)$ has an analytic extension and
the functions $\O_1(k),\O(k)$ have meromorphic extensions
from $\K_1\sm\s_d$
onto the whole complex plane.

\begin{lemma} Let $p,q\in\cH_0$, and let $k\in\C\sm\{0\}$. Then
the determinant $D$ satisfies
\[
\lb{cS(k)}
D(ik)=D(k)\det S(k),
\]
\[
\lb{cT(k)}
D(-k)=D(k)\det\O(k).
\]
The function $\det S(k)$ is continuous in $\R_+$.

\end{lemma}

\no {\bf Proof.}
The identities \er{defP}, \er{cP1}  and \er{2.5} give
$$
\begin{aligned}
D(ik)=\det\big(1+Y_0(ik)\big)=\det\big(1+Y_0(k)+P_1(k)\big)
=D(k)\det\Big(1+\big(1-Y(k)\big)P_1(k)\Big)
\\
=D(k)\det\Big(1+c_k\big(1-Y(k)\big)\p_2(k)\p_1(k)\Big).
\end{aligned}
$$
The definitions \er{S}, \er{ScA} give
$$
S(k)=1+c_k\p_1(k)\big(1-Y(k)\big)\p_2(k).
$$
The identity \er{2.2} implies \er{cS(k)}.

Similarly, the identities \er{defP}, \er{cP2}  and \er{2.5}  give
$$
\begin{aligned}
D(-k)=\det\big(1+Y_0(-k)\big)
=\det\big(1+Y_0(k)+P_2(k)\big)
=D(k)\det\Big(1+\big(1-Y(k)\big)P_2(k)\Big)
\\
=D(k)\det\Big(1+c_k\big(1-Y(k)\big) \P_2(k)\P_1(k)\Big).
\end{aligned}
$$
Then
the identity \er{2.2} and the definition
\er{defOmega} imply \er{cT(k)}.

Due to Lemma~\ref{TA}, the function $\det S(k)$ is continuous in
$k\in \R_+\sm\s_d$ and it has a
meromorphic extension from $\R_+$ onto $\C$.
Moreover, if $k\in\s_d\cap\R_+$,
then $k$ is a zero of the functions $D(ik)$
and $D(k)$ of the same multiplicity. Due to the identity
\er{cS(k)}, $\det S(k)$ is continuous
at the point $k\in\s_d$.
Therefore, $\det S(k)$ is continuous in $\R_+$.
\BBox

\subsection{Asymptotics of $\O$}
We consider the matrix-valued function
$\O=\1_4+c_k\big(\O_0-\O_1\big)$. Substituting the definitions
\er{defPsi12} into \er{defOm0}, \er{defOm11} we obtain
\[
\lb{idOm01}
\begin{aligned}
\O_0(k)=\P_1(k)\P_2(k)
=\ma\p_1(ik)\\ \p_1(k)\am\Big(i\p_2(ik), \p_2(k)\Big),
\\
\O_1(k)=\P_1(k)Y(k)\P_2(k)
=\ma\p_1(ik)\\ \p_1(k)\am Y(k)\Big(i\p_2(ik), \p_2(k)\Big).
\end{aligned}
\]
Introduce entire $2\ts 2$ matrix-valued functions
\[
\lb{defmB}
\mB_1(k)=\p_1(ik)\p_2(k),
\qq
\mB_2(k)=i\p_1(k)\p_2(ik),
\]
The identities \er{idOm01}
and the definitions \er{ScA01}
\er{defmB} give
\[
\lb{O0cAmB}
\O_0(k)=\ma i\cA_0(ik)&\mB_1(k)&\\\mB_2(k)&\cA_0(k)\am,
\]
\[
\lb{defOm1}
\O_1(k)
=\ma i\p_1(ik)Y(k)\p_2(ik)& \p_1(ik)Y(k)\p_2(k)\\
i\p_1(k)Y(k)\p_2(ik)&\cA_1(k)\am.
\]
Introduce the domain
$$
\K_1^+=\Big\{k\in\K_1:\arg k\in\Big(0,{\pi\/4}\Big)\Big\}.
$$

\begin{lemma}
Let $p,q\in\cH_0$. Then
the functions
$\mB_1(k),\mB_2(k)$, given by \er{defmB}, satisfy
\[
\lb{defmB1}
\begin{aligned}
\mB_1(k)={1\/2\pi}\ma\b_1(k)&\b_2(k)\\\b_2(-k)&\b_1(-k)\am,
\\
\mB_2(k)={1\/2\pi}\ma i\b_1(-k)&i\b_2(k)\\i\b_2(-k)&i\b_1(k)\am,
\end{aligned}
\]
where
\[
\lb{beta12}
\b_1(k)=\int_\R\big(q(x)-i2k^2p(x)\big)e^{(1+i)kx}dx,\qq
\b_2(k)=\int_\R\big(q(x)+i2k^2p(x)\big)e^{(1-i)kx}dx.
\]
Moreover, the function $\det\O(k)$ satisfies
\[
\lb{asdetOm}
\det\O(k)=e^{2\g(\Re k+\Im k)}O(k^{-1}),
\]
as $|k|\to\iy, k\in\ol\K_1^+$, uniformly in $\arg k\in[0,{\pi\/4}]$.

\end{lemma}

\no {\bf Proof.}
Substituting the definitions \er{defPsi}, \er{dpxk}
into \er{defmB} we obtain the identities \er{defmB1}.

 Let $k\in\ol\K_1^+,|k|\to\iy$.
The definitions \er{beta12} give
$$
\begin{aligned}
\b_1(k)=e^{(\Re k-\Im k)\g}O(k^2),&\qq
\b_1(-k)=O(k^2),\\
\b_2(k)=e^{(\Re k+\Im k)\g}O(k^2),&\qq
\b_2(-k)=O(k^2).
\end{aligned}
$$
Substituting these asymptotics into the identities \er{defmB1}
we obtain
$$
\begin{aligned}
\mB_1(k)=\ma e^{(\Re k-\Im k)\g}O(k^2)&e^{(\Re k+\Im k)\g}O(k^2)\\
O(k^2)&O(k^2)\am,
\\
\mB_2(k)=\ma O(k^2)&e^{(\Re k+\Im k)\g}O(k^2)\\
O(k^2)&e^{(\Re k-\Im k)\g}O(k^2)\am.
\end{aligned}
$$
Substituting these asymptotics and \er{asA0r}
into the identity \er{defOm0} we obtain
\[
\lb{asOm0pr}
\O_0(k)=
\ma O(k^2)&e^{2\g\Re k}o(k^2)&
e^{(\Re k-\Im k)\g}O(k^2)&e^{(\Re k+\Im k)\g}O(k^2)\\
o(k^2)&O(k^2)&O(k^2)&O(k^2)\\
O(k^2)&e^{(\Re k+\Im k)\g}O(k^2)
 &O(k^2)&e^{2\g\Im k}o(k^2)\\
O(k^2)&e^{(\Re k-\Im k)\g}O(k^2)&o(k^2)&O(k^2)\am.
\]

Let $|k|\to\iy, k\in\ol\K_1$.
The estimates \er{Pe} and \er{Y2} give
$$
\begin{aligned}
\p_1(ik)Y(k)\p_2(ik)=\ma e^{\g\Re k}O(k)&e^{2\g\Re k}O(k)\\
O(k)&e^{\g\Re k}O(k)\am,
\\
\p_1(ik)Y(k)\p_2(k)=\ma e^{\g\Re k}O(k)&e^{\g(\Re k+\Im k)}O(k)\\
O(k)&e^{\g\Im k}O(k)\am,
\\
\p_1(k)Y(k)\p_2(ik)=\ma e^{\g\Im k}O(k)&e^{\g(\Re k+\Im k)}O(k)\\
O(k)&e^{\g\Re k}O(k)\am.
\end{aligned}
$$
Substituting these asymptotics and \er{asA1xim}
into the identity \er{defOm1} we obtain
\[
\lb{asOm1pr}
\O_1(k)=
\ma e^{\g\Re k}O(k)&e^{2\g\Re k}O(k)&
e^{\g\Re k}O(k)&e^{\g(\Re k+\Im k)}O(k)\\
O(k)&e^{\g\Re k}O(k)&O(k)&e^{\g\Im k}O(k)\\
e^{\g\Im k}O(k)&e^{\g(\Re k+\Im k)}O(k)
&e^{\g\Im k}O(k)&e^{2\g\Im k}O(k)\\
O(k)&e^{\g\Re k}O(k)&O(k)&e^{\g\Im k}O(k)\am.
\]

Substituting the asymptotics \er{asOm0pr}, \er{asOm1pr} into
the definition \er{defPsi} we obtain the asymptotics
$$
\O(k)=
\ma e^{\g\Re k}O(k^{-1})&e^{2\g\Re k}O(k^{-1})&e^{\g\Re k}O(k^{-1})&e^{\g(\Re k+\Im k)}O(k^{-1})\\
O(k^{-1})&e^{\g\Re k}O(k^{-1})&O(k^{-1})&e^{\g\Im k}O(k^{-1})\\
e^{\g\Im k}O(k^{-1})&e^{\g(\Re k+\Im k)}O(k^{-1})&1+e^{\g\Im k}O(k^{-1})&e^{2\g\Im k}O(k^{-1})\\
O(k^{-1})&e^{\g\Re k}O(k^{-1})&O(k^{-1})&1+e^{\g\Im k}O(k^{-1})\am,
$$
as $|k|\to\iy, k\in\ol\K_1^+$,
which yields \er{asdetOm}.
\BBox

\section{Proof of the main Theorems}
\setcounter{equation}{0}

\subsection{Asymptotics of the determinant}
We prove our preliminary  Proposition \ref{T1}.

\medskip

\no {\bf Proof of Proposition \ref{T1}.}
i) The statement is proved in Lemma~\ref{TY}~i).

ii) Due to Lemma~\ref{TD1}, the function $D(k)$ has an
analytic extension from $\K_1$ onto
$\C\sm\{0\}$,
it is real on the line
$e^{i{\pi\/4}}\R$ and the function $k^4D(k)$ is entire.
The asymptotics \er{aD1} yields the asymptotics \er{asDK+}.
This asymptotics shows that the function $D(k)$ has
a finite number of zeros in $\ol\K_1$. Then
the operator $H$ has a finite number of eigenvalues.
\BBox

\medskip

We determine asymptotics of the determinant
in the complex plane. Due to the symmetry of $D(k)$
we need to get this asymptotics in the domains
$\K_1,\K_2,\K_3$.
The asymptotics in the domain $\K_1$
is known due to \er{aD1}.
We analyze the function $D(k)$ in the domains $\K_2,\K_3$
by the following way.
We obtain the asymptotics
of $S(k)$ and $\O(k)$ in $\K_1$. Then
we use the identities \er{cS(k)}, \er{cT(k)}
in order to determine the asymptotics of $D(ik),D(-k)$
in $\K_1$, which gives the asymptotics of $D(k)$
in $\K_2,\K_3$.

\begin{lemma} Let $p,q\in\cH_0$. Then
\[
\lb{asSf}
\det S(k)=1+O(k^{-1})+e^{2\g\Im k}O(k^{-2}),
\]
\[
\lb{estDr}
D(ik)=1+O(k^{-1})+e^{2\g\Im k}O(k^{-2}),
\]
as $k\in\ol\K_1,|k|\to\iy$, uniformly in
$\arg k\in[0,{\pi\/2}]$,
\[
\lb{estDr2}
D(-k)=1+e^{2\g(\Re k+\Im k)}O(k^{-1})
\]
as $k\in\ol\K_1^+,|k|\to\iy$, uniformly in
$\arg k\in[0,{\pi\/4}]$.
\end{lemma}

\no {\bf Proof.}
Let $|k|\to\iy, k\in\ol\K_1$.
Substituting the asymptotics \er{asA0r} and
\er{asA1xim} into
the identity \er{S} we obtain the asymptotics
$$
S(k)=\ma 1+O(k^{-1})+e^{\g\Im k}O(k^{-2})&e^{2\g\Im k}o(k^{-1})\\
o(k^{-1})&1+O(k^{-1})+e^{\g\Im k}O(k^{-2})\am,
$$
which yields the asymptotics \er{asSf}.
Substituting the asymptotics \er{asDK+} and \er{asSf}  into
\er{cS(k)} we obtain
the asymptotics \er{estDr}.

Substituting the asymptotics \er{asDK+},
\er{asdetOm} into the identity \er{cT(k)} we obtain
\er{estDr2}.
\BBox

\medskip

We prove Theorem~\ref{CorEstN}.

\medskip

\no {\bf Proof of Theorem~\ref{CorEstN}}.
The asymptotics \er{asDK+} gives the estimate \er{unifestD} in $\K_1$,
the asymptotics \er{estDr} gives \er{unifestD} in $\K_2$,
the asymptotics \er{estDr2} gives \er{unifestD} in $\K_3$.
The estimate \er{unifestD} in $\K_2$ and the symmetry
$D(k)=\ol{D(i\bar k)}$ imply the estimate
\er{unifestD} in $\K_4$ .

The asymptotics \er{estDr} yields $|k(D(k)-1)|\le Ce^{-2\g\Re k}$
for all $k\in\K_2$ for some $C>0$. Let $k_*\in\K_2$ be a
resonance. Then the identity $D(k_*)=0$ gives the estimate
\er{estFD}.

We prove the estimate \er{estnr}.
Recall that the function $D(k)$ is analytic in $\C\sm\{0\}$
and may have a pole of order $\le 4$ at the point $k=0$.
Let the function $F(k)=k^mD(k),m\le 4$, be entire
and satisfy $F(0)\ne 0$.
Let $\cN_F(r)$ be the number of zeros
of the function $F$ in the disc $|k|<r$ counted with multiplicity.
If $D(0)\ne 0$, then $\cN=\cN_F$,
if $k=0$ is a zero of $D$
of multiplicity $\ell$, then $\cN=\cN_F+\ell$.
We have to prove that $\cN_F$ satisfies
the estimate \er{estnr}.
The estimate \er{unifestD} gives
\[
\lb{esF}
\log|F(k)|\le2\g \big((\Re k)_-+(\Im k)_-\big)+
C\log |k|
\]
for all $k\in\C,|k|$ large enough and
for some $C>0$.
Substituting the estimate \er{esF} into Jensen's formula
\[
\lb{JFg}
\int_0^r\fr{\cN_F(t)}{t}dt=
\fr{1}{2\pi}\int_0^{2\pi}\log |F(re^{i\theta})|d\theta
-\log|F(0)|,
\]
we obtain
$$
\int_0^r\fr{\cN_F(t)}{t}dt\le
-\fr{\g r}{\pi}\Big(
\int_{\fr{\pi}{2}}^{\pi}\cos\theta d\theta
+\int_{\pi}^{\fr{3\pi}{2}}
(\cos\theta+\sin\theta ) d\theta
+\int_{\fr{3\pi}{2}}^{2\pi}\sin\theta d\theta\Big)
+ C\log r=
\fr{4\g r}{\pi}+ C\log r
$$
for all $r>0$ large enough.
Then there exists
$$
\lim_{r\to+\iy} {1\/r}\int_0^r\fr{\cN_F(t)}{t}dt\le{4\g\/\pi}.
$$
The estimate \er{estnr} follows from the following
well known result,
see, e.g., \cite[Lm II.4.3]{Le96}:

{\it
Let $\cN_F(t)$ be non-decreasing function on $\R_+$,
$\cN_F(t)=0$ as $0\le t<\ve$ for some $\ve>0$, and let the function
$$
I(r)=\fr{1}{r}\int_0^r\fr{\cN_F(t)}{t}dt,\qqq r\in\R_+,
$$
has the limit as $r\to\iy$.
Then
$
\cN_F(r)=r(I(r)+o(1))
$
as $r\to\iy$.
}
\BBox

\subsection{Trace formulas}
Let $\z_n,n\in\N$, be the zeros
of the function $D$ in $\C\sm\{0\}$
labeled by $0<|\z_1|\le|\z_2|\le...$  counting with multiplicities.
The estimate \er{unifestD} provides
the standard Hadamard factorization
\[
\lb{HadF}
D(k)={\a\/k^{m}} e^{\b k}
\lim_{r\to\iy}
\prod_{|\z_n|<r}\Big(1-{k\/\z_n}\Big)e^{k\/\z_n},\qqq
m\le 4,
\]
absolutely and uniformly on any compact
subset in $\C\sm\{0\}$, where $\a,\b$ are defined in \er{Dat0}.
The identity \er{HadF} gives
\[
\lb{expF'1}
{D'(k)\/D(k)}=\b-{m\/k}+k\lim_{r\to\iy}\sum_{|\z_n|<r}
{1\/\z_n (k-\z_n)}.
\]
The following proofs use the approach from \cite{K04}.

\medskip

\no {\bf Proof of Theorem \ref{ThRes}.}
Let $k\in\K_1\sm\s_d$.
The definitions \er{defVj}, \er{R01} show that the operators
$VR_0(k),V_2R_0(k)$ and $R_0(k)V_1$
are Hilbert-Schmidt. Then the operator
$$
R_0(k)-R(k)=R_0(k)VR(k)=R_0(k)VR_0(k)-R_0(k)VR_0(k)VR(k)
$$
is trace class.
Due to the identities \er{2.2}, \er{2.3}, \er{2.5} and
$Y_0'(k)=4k^3V_2R_0^2(k)V_1$,
the derivative of $D$ satisfies
\[
\lb{idTrRR0}
{1\/4k^3}{D'(k)\/D(k)}=\Tr \big((I+Y_0(k))^{-1}V_2R_0^2(k)V_1\big)=
\Tr R_0(k)VR(k)=\Tr (R_0(k)-R(k)).
\]
The identity \er{expF'1} together with \er{idTrRR0} yields
the trace formula \er{expF'}.
\BBox

\medskip

The S-matrix $S(k), k\in\R_+$, is a complex $2\ts 2$
matrix and
$|\det S(k)|=1$. Thus we have
\[
 \label{Sx}
 \det S(k)=e^{-2\pi i\f_{sc}(k)},\qqq k\in\R_+.
\]
Since
$\det S(k)$ is continuous in $\R_+$
and $\det S(k)=1+O(k^{-1})$ as $k\to +\iy$
(see \er{asSf}), formula \er{Sx} determines
$\f_{sc}(k)$ by the identity $\f_{sc}(k)={i\/2\pi }\log\det S(k)$,
the continuity, and the
asymptotics $\f_{sc}(k)=O(k^{-1})$ as $k\to +\iy$.

\begin{theorem}
\lb{ThScATr}
Let
$(p,q)\in\cH_0$.
Then  \[
\lb{expphi'}
\phi_{sc}'(k)
={1\/2\pi i}\lt((1-i)\b
+\sum_{n=1}^\iy{k\/\z_n}\Big({1\/ik-\z_n}
+{1\/k-\z_n}\Big)\rt),\qq k\in\R_+\sm\s_d,
\]
the series converges absolutely and uniformly on any compact subset in
$\R_+\sm\s_d$.
\end{theorem}

\no {\bf Proof.}
The function $\det S(k)$ is continuous in $\R_+$,
has a meromorphic extension onto the whole complex plane
and, due to equations \er{Sx} and \er{cS(k)},
it satisfies the identities
$$
e^{-2\pi i\phi_{sc}(k)}=\det S(k)={D(ik)\/D(k)},
\qqq\forall\qq k>0.
$$
Differentiating  this identity
we obtain
$$
-2\pi i\phi_{sc}'(k)
=i{D'(ik)\/D(ik)}-{D'(k)\/D(k)}.
$$
Then the identity \er{expF'1} implies \er{expphi'}.
\BBox

\subsection{The Euler-Bernoulli operator.}
\lb{sect55}
We consider the Euler-Bernoulli operator
\[
\lb{defcE}
\cE u={1\/b}(au'')'', \qqq
\]
acting on $L^2(\R,b(x)dx)$,
where the  coefficients $a,b$
satisfy
\[
\lb{abc}
a>0,\qq b>0,\qq
a-1,b-1\in\cH_4(1).
\]

Now we consider the Liouville type transformation of the operator
$\cE$ into the operator $H$, defined by \er{a.1}
with specific $p,q$ depending on $a,b$. In
order to define this transformation we introduce the new variable
$t\in\R$ by
\[
\lb{tx}
t=t(x)=\int_{0}^x\Big({b(s)\/a(s)}\Big)^{1\/4}ds,\qqq\forall\qq x\in\R.
\]
Let $x=x(t)$ be the inverse
function for $t(x), x\in\R$. Introduce the unitary
transformation
$U:L^2(\R,b(x)dx) \to L^2(\R,dt)$ by
\[
\lb{defU}
u(x)\mapsto y(t)=(Uu)(t)=a^{1\/8}(x(t))b^{3\/8}(x(t))u(x(t))
\qqq\forall\qq t\in\R.
\]

Introduce the functions $\a(t),\b(t),t\in\R$, by
\[
\lb{abxe}
\a(t)={1\/a(x(t))}{da(x(t))\/dt},\qqq \b(t)={1\/b(x(t))}{db(x(t))\/dt}.
\]
Then the functions $\a,\b\in L^1(\R)$ are real, compactly supported
and satisfy
$$
\stackrel{\ldots}{a},\stackrel{\ldots}{\b}\in L^1(\R),\qqq
\where\qq \dot f={df\/dt}.
$$

Let the
operator $H$ be defined by \er{a.1},
where the coefficients $p(t),q(t),t\in\R$, have the forms
\[
\lb{peb}
p=-{\dot\e_0+\vk\/2},
\]
\[
\lb{qeb}
q={d\/dt}\big((\dot \e_2+\e_2^2)\e_1-\ddot \e_1\big)
+\big((\dot \e_2+\e_2^2)\e_1-\ddot \e_1\big)\e_1,
\]
and the functions $\vk(t),\e_0(t),\e_1(t),\e_2(t)$ are given by
\[
\lb{kapx}
\vk={5\a^2+5\b^2+6\a\b\/32}\ge {\a^2+\b^2\/16}\ge 0,
\]
$$
\e_0={3\a+5\b\/4},\qqq
\e_1={\a+3\b\/8},\qqq \e_2={3\a+\b\/8}.
$$
The coefficients satisfy:
$p,p'',q\in L^1(\R_+),(p,q)\in\cH_2\ts\cH_0$
with $\g$ given by \er{gammaEB}.

Let the coefficients $a,b$  satisfy the conditions
\er{abc}. Let the operator $\cE$ be defined by \er{defcE}
and let the operator $H$ be defined by \er{a.1},
where the coefficients
$p,q$ have the forms \er{peb}, \er{qeb}.
Repeating the arguments from \cite{BK15} we obtain that
the operators $\cE$ and $H$ are unitarily equivalent and
satisfy:
\[
\lb{eqEH}
\cE=U^{-1}HU.
\]
where the operator $U$ is defined by \er{defU}.

\begin{corollary}
Let $a-1,b-1\in\cH_4(1)$ and let $a,b$ be positive.
Then the determinant $D(k)$ satisfies
\[
\lb{asdetGEB}
D(k)=1+\fr{1+i}{4k}\int_\R\vk(t) dt+{O(1)\/k^2}\qqq
\as \qqq |k| \to \infty, \qq  k\in\ol\K_1
\]
uniformly in $\arg k\in[0,{\pi\/2}]$, where $\vk(t)$
is given by the definition \er{kapx}.
\end{corollary}

\no {\bf Proof.}
Identity \er{peb} gives
$$
p_0=\int_\R p(t)dt=-{1\/2}\int_\R\vk(t) dt.
$$
Substituting this identities into the asymptotics
\er{asDK+} we obtain the asymptotics \er{asdetGEB}.
\BBox

\medskip

The definition \er{kapx} shows that $\vk\ge 0$, moreover, $\vk=0$
iff $\a=\b=0$. Then the second term in the asymptotics
\er{asdetGEB} vanishes iff $\a=\b=0$.
The proof of Theorem \ref{ThEB} is based on
this observation.

\medskip

\no {\bf Proof of Theorem \ref{ThEB}.}
Assume that the operator $\cE$ does not have any eigenvalues and
resonances. Then $D=1$ and the second term in the asymptotics
\er{asdetGEB} vanishes. The estimates \er{kapx} show that $\a=\b=0$
in this case, then  $a=b=1$ on $\R$.

Conversely,
assume that $a=b=1$ on $\R$. Then $\a=\b=0$ and
the definitions \er{peb}, \er{qeb} imply
$p=q=0$. The identities \er{defVj} yield
$V_1=0,V_2=0$. Then the definition \er{Y0}
gives $Y_0=0$, and the identity \er{a.2}
implies $D=1$. Therefore,
there are not any eigenvalues and resonances.
\BBox

\section{Asymptotics of the resonances}
\setcounter{equation}{0}

\subsection{Asymptotics of the determinant}
The function $D(k)$ has a finite number of zeros in the domain
$\K_1$. The identity \er{cS(k)} shows that $ik$
with large $|k|$ is a resonance in $\K_2$ iff
$k$ is a zero of the function $\det S(k)$ in $\K_1$.
Thus in order to determine asymptotics of resonances
in $\K_2$ we need to improve asymptotics of $\det S(k)$ in $\K_1$.
Similarly, the identity \er{cT(k)} shows that $-k$
with large $|k|$ is a resonance in $\K_3$ iff
$k$ is a zero of the function $\det\O(k)$ in $\K_1$.
Then in order to determine asymptotics of resonances
in $\K_3$ we have to improve asymptotics of the
function $\det\O(k)$ in $\K_1$. Moreover,
due to the symmetry of the determinant it is sufficiently
to consider in this case the domain
$$
\K_1^+=\big\{k\in\C:\arg k\in\big(0,\tf{\pi}{4}\big)\big\}.
$$

\begin{lemma}
\lb{LmAsSOm}
Let $p,q$ satisfy the conditions
\er{ppm}.
Then the S-matrix $S(k)$ and the matrix-valued function
$\O(k)$, defined by \er{defOmega}, satisfy
\[
\lb{asS1}
\det S(k)=1-{e^{-i2k\g}p_+p_-\/(2k)^4}\big(1+o(1)+e^{-\g\Im k}O(k)\big)+O(k^{-1}),
\]
as $|k|\to\iy, k\in\ol\K_1$, uniformly in $\arg k\in[0,{\pi\/2}]$,
\[
\lb{ascT}
\det\O(k)={e^{2(1-i)k\g}(p_+p_-)^2\/(2k)^8}\big(1+o(1)+e^{-\g\Im k}O(k)\big)
-{e^{2k\g}p_+p_-\/(2k)^4}\big(1+o(1)\big),
\]
as $|k|\to\iy,k\in \ol \K_1^+$,
uniformly in $\arg k\in[0,{\pi\/4}]$.

\end{lemma}

\no {\bf Proof.}
Let $k\in\ol\K_1,|k|\to\iy$. The definition \er{S} and
the asymptotics \er{asA0} and \er{ascA1im} give
\[
\lb{idSord}
S(k)=\1_2+c_k\big(\cA_0(k)-\cA_1(k)\big)={1\/4k^2}E_1^+(k)G_1(k)E_1^-(k),
\]
where
\[
\lb{h2pm}
c_k={\pi\/i2k^3},\qq
E_1^+(k)=\ma e^{-i\g k}&0\\0&1\am,\qq
E_1^-(k)=\ma 1&0\\0&e^{-i\g k}\am,
\]
\[
\lb{asDor}
G_1(k)=\ma 4k^2e^{ik\g}\big(1+O(k^{-1})\big)+O(k^{-1})&p_++o(1)+e^{-2\g\Im k}O(1)\\
p_-+o(1)+e^{-2\g\Im k}O(1)&4k^2e^{ik\g}\big(1+O(k^{-1})\big)+O(k^{-1})\am.
\]
The asymptotics \er{asDor} implies
$$
\det G_1(k)=16k^4e^{2ik\g}\big(1+O(k^{-1})\big)-p_+p_-+o(1)+e^{-\g\Im k}O(k).
$$
The identity \er{idSord} and the definitions \er{h2pm} yield
the asymptotics \er{asS1}.

We prove the asymptotics \er{ascT}.
Substituting the identities
\er{O0cAmB}, \er{defOm1} into the definition \er{defOmega}
and using the definition \er{S}
we obtain
\[
\lb{idOmega}
\O(k)=\ma \1_2+c_k\big(i\cA_0(ik)-i\p_1(ik)Y(k)\p_2(ik)\big)
&c_k\big(\mB_1(k)-\p_1(ik)Y(k)\p_2(k)\big)
\\c_k\big(\mB_2(k)-i\p_1(k)Y(k)\p_2(ik)\big)
&S(k)\am.
\]
Let $k\in\ol\K_1^+,|k|\to\iy$ and let $p_+p_-\ne 0$.
Integrations by parts in the definitions \er{beta12}
give
$$
\begin{aligned}
\b_1(k)
=e^{(\Re k-\Im k)\g}O(k),\qqq
\b_1(-k)
=O(k),\\
\b_2(k)=-(1-i)ke^{(1-i)k\g}(p_++o(1)),\qq
\b_2(-k)=-(1-i)k(p_-+o(1)).
\end{aligned}
$$
Substituting these asymptotics into the identities \er{defmB1}
we obtain
\[
\lb{asmB12}
\mB_1(k)=-{k\/2\pi}\ma e^{(\Re k-\Im k)\g}O(1)
&(1-i)e^{(1-i)k\g}(p_++o(1))\\
(1-i)(p_-+o(1))&O(1)\am,
\]
\[
\lb{asmB122}
\mB_2(k)=-{k\/2\pi}\ma O(1)
&(1+i)e^{(1-i)k\g}(p_++o(1))\\
(1+i)p_-+o(1)&e^{(\Re k-\Im k)\g}O(1)\am.
\]
Repeating the arguments from the proof of the asymptotics
\er{ascA1im} we obtain
\[
\lb{pYp1}
\p_1(ik)Y(k)\p_2(ik)=\ma e^{\g\Re k}O(1)&e^{2\g\Re k}O(1)\\
O(1)&e^{\g\Re k}O(1)\am,
\]
\[
\lb{pYp2}
\p_1(ik)Y(k)\p_2(k)=\ma e^{\g\Re k}O(1)&e^{\g(\Re k+\Im k)}O(1)\\
O(1)&e^{\g\Im k}O(1)\am,
\]
\[
\lb{pYp3}
\p_1(k)Y(k)\p_2(ik)=\ma e^{\g\Im k}O(1)&e^{\g(\Re k+\Im k)}O(1)\\
O(1)&e^{\g\Re k}O(1)\am.
\]
The asymptotics \er{ascA0ik} and \er{pYp1} give
\[
\lb{idAord}
\1_2+c_k\big(i\cA_0(ik)-i\p_1(ik)Y(k)\p_2(ik)={1\/4k^2}E_2^+(k)G_2(k)E_2^-(k),
\]
where
\[
\lb{h1pm}
E_2^+(k)=\ma e^{\g k}&0\\0&1\am,\qq
E_2^-(k)=\ma 1&0\\0&e^{\g k}\am,
\]
\[
\lb{defAord}
G_2(k)=\ma O(k^{-1})&-p_++o(1)\\-p_-+o(1)&O(k^{-1})\am.
\]
The asymptotics \er{asmB12} and \er{pYp2} imply
\[
\lb{idBord}
c_k\big(\mB_1(k)-\p_1(ik)Y(k)\p_2(k)\big)={1\/4k^2}E_2^+(k)G_3(k)E_1^-(k),
\]
where
\[
\lb{asBor}
G_3(k)=\ma O(k^{-1})+e^{-\g\Im k}O(1)&(1+i)p_++o(1)\\
(1+i)p_-+o(1)&O(k^{-1})+e^{-\g\Im k}O(1)\am.
\]
The asymptotics \er{asmB122} and \er{pYp3} yield
\[
\lb{idCord}
c_k\big(\mB_2(k)-i\p_1(k)Y(k)\p_2(ik)\big)={1\/4k^2}E_1^+(k)G_4(k)E_2^-(k),
\]
where
\[
\lb{asCor}
G_4(k)=\ma O(k^{-1})+e^{-\g\Im k}O(1)&i(1+i)p_+(1+o(1))\\
i(1+i)p_-(1+o(1))&O(k^{-1})+e^{-\g\Im k}O(1)\am.
\]
Substituting the identities \er{idSord}, \er{idAord}, \er{idBord} and
\er{idCord} into the relation \er{idOmega} we obtain the identity
\[
\lb{idOm1}
\O(k)={1\/4k^2}\ma E_2^+(k)&0\\0&E_1^+(k)\am
\ma G_2(k)&G_3(k)\\ G_4(k)&G_1(k)\am\ma E_2^-(k)&0\\0&E_1^-(k)\am.
\]
The definitions \er{h2pm}, \er{h1pm} and the identity \er{idOm1} yield
\[
\lb{idetwOm1}
\det\O(k)={e^{2(1-i)k\g}\/(2k)^8}\det\ma G_2(k)&G_3(k)\\ G_4(k)&G_1(k)\am.
\]
The standard matrix formula gives
$$
\det\ma G_2(k)&G_3(k)\\ G_4(k)&G_1(k)\am=\det G_2(k)\det G(k),
$$
where
$$
G(k)=G_1(k)-G_4(k)G_2^{-1}(k)G_3(k).
$$
Substituting this identity into \er{idetwOm1} we obtain
\[
\lb{idetwOm}
\det\O(k)={e^{2(1-i)k\g}\/(2k)^8}\det G_2(k)\det G(k).
\]
The definition \er{defAord} implies
\[
\lb{asdetA}
\det G_2(k)=-p_+p_-\big(1+o(1)\big),\qq
G_2^{-1}(k)=\ma O(k^{-1})&-{1\/p_-}(1+o(1))\\-{1\/p_+}(1+o(1))&O(k^{-1})\am.
\]
The asymptotics \er{asDor}, \er{asBor}, \er{asCor}  and \er{asdetA} yield
$$
G_4(k)G_2^{-1}(k)=\ma -i(1+i)+o(1)&
e^{-\g\Im k}O(1)+O(k^{-1})\\
e^{-\g\Im k}O(1)+O(k^{-1})&-i(1+i)+o(1)\am,
$$
$$
G_4(k)G_2^{-1}(k)G_3(k)=\ma e^{-\g\Im k}O(1)+O(k^{-1})&2p_++o(1)+e^{-2\g\Im k}O(1)\\
2p_-+o(1)+e^{-2\g\Im k}O(1)&e^{-\g\Im k}O(1)+O(k^{-1})\am,
$$
$$
G(k)=\ma 4k^2e^{ik\g}\big(1+O(k^{-1})\big)+O(k^{-1})&
-p_++o(1)+e^{-2\g\Im k}O(1)\\
-p_-+o(1)+e^{-2\g\Im k}O(1)&
4k^2e^{ik\g}\big(1+O(k^{-1})\big)+O(k^{-1})\am,
$$
which yields
\[
\lb{asD-CAB}
\det G(k)=
16k^4e^{2ik\g}\big(1+O(k^{-1})\big)-p_+p_-+o(1)+e^{-\g\Im k}O(k).
\]
Substituting the asymptotics \er{asdetA} and \er{asD-CAB}
into the identity \er{idetwOm} we obtain
the asymptotics
\er{ascT}.
\BBox

\subsection{Asymptotics of resonances}
We are ready to determine asymptotics of resonances.

\medskip

\no {\bf Proof of Theorem \ref{ThAsCF}.}
Let $k\in\K_1,|k|\to\iy$ and let $ik$ be a resonance.
The identity \er{cS(k)} shows that
$k$ is a zero of the function $\det S(k)$ in $\K_1$.
The asymptotics \er{asS1} and
the identity
$\det S(k)=0$ imply that $k$ satisfies
the equation
$$
k^4={p_+p_-\/16}e^{-i2k\g}
\big(1+o(1)+e^{-\g\Im k}O(k)\big).
$$
Then $k$ lies on the logarithmic curve $\G$ in $\K_1$, given by
$$
|k|={|p_+p_-|^{1\/4}\/2}e^{{1\/2}\g\Im k}\big(1+o(1)\big),
$$
and satisfies
\[
\lb{asriK+}
ik={ij_n\pi\/\g}-{2\log k\/\g}+{1\/2\g}\log {|p_+p_-|\/16}+o(1)
\]
and there are not any other large resonances in $i\K_+$.

Let $k\in \K_1^+$, let $-k$ be a resonance and let $|k|$ be large
enough.
The identity \er{cT(k)} shows that $-k$
is a zero of the function $\det\O(k)$ in $\K_1$.
The identity $\det\O(k)=0$ and the asymptotics \er{ascT} gives
$$
k^4={p_+p_-\/16}e^{-2ik\g}\big(1+o(1)+e^{-\g\Im k}O(k)\big).
$$
Then $k$ lies on the curve $\G$
and satisfies
\[
\lb{asriK-}
-k=-{j_n\pi\/\g}-{i2\log k\/\g}+{i\/2\g}\log {|p_+p_-|\/16}+o(1)
\]
and there are not any other large resonances in $-\K_1^+$.
The asymptotics \er{asriK+}, \er{asriK-} give
the asymptotics \er{asresiK+}, which yields
the asymptotics \er{asNj}.

The estimates \er{unifestD} and \er{Dat0} yield
that the determinants
$k^4D(k)$ is of exponential type. But the asymptotics \er{asresiK+}
and the asymptotics of the Levinson Theorem \er{AsN}  show that the
determinants $D(k)$ is not of the Cartwright class. \BBox

\subsection{Further discussions}
\lb{SectFD}

We will discuss what properties of resonances of the second and
fourth order operators with compactly supported coefficients are common and
which are specific.

$\bu $ Common properties:

1) The determinants $D(k)$ and $d(k)$ are exponentially type
functions of the variable $k$ and each of them
has an axis of symmetry.

2)  For coefficients with steps
the resonances have the logarithmic type asymptotics.

$\bu $ Specific  properties of the determinant $d(k)$
for a Schr\"odinger operator:

1)  In terms of the spectral parameter $\l$
the Riemann surface for the determinant $d(\l^{1\/2})$
 is the two sheeted
Riemann surface for the function $\l^{1\/2}$.
The function $d(k)\sim 1$ as $|\l|\to \iy$ on
the physical sheet and $d(k)\sim e^{2\g|\Im \sqrt \l|}$ on
the non-physical sheet. It has a finite number of zeros
(eigenvalues) on the physical sheet and an infinite number of
zeros (resonances)
on the non-physical one.

2) The determinant $d(k)$ belongs to the Cartwright class $\cC_{Cart}$.
Then the Levinson Theorem gives the distribution of resonances
in the large disc.

3) The number of resonances in the disk $|\l|<r$
for large $r$ has
asymptotics ${2\g \/\pi}r^{1\/2}(1+o(1))$.

4) Using one identity \er{SD2od} we
obtain an analytic extension of the determinant from
the physical sheet onto the non-physical one.

$\bu $ Specific  properties of the determinant $D(k)$  for a fourth
order operator:

1)  The Riemann surface for the determinant $D(\l^{1\/4})$ is the four
sheeted
Riemann surface for the function $\l^{1\/4}$.
The function $D$ satisfies:
$D\sim 1$ at $|\l|\to \iy$ on the first sheet, $D\sim
e^{2\g|\l|^{1\/4}}$ on the second and fourth sheets and $D\sim
e^{2\sqrt2\g|\l|^{1\/4}}$ on the third sheet. It has a finite
number of zeros (eigenvalues) on the first (physical) sheet and an infinite
number of zeros (resonances) on the other (non-physical) sheets. The number
of resonances in the large disc on the third sheet is, roughly
speaking, in two times more than on the second (or fourth) sheet.

2) The determinant $D$ is not in the Cartwright class.

3) The number of resonances in the disk $|\l|<r$ has
asymptotics ${4\g\/\pi}r^{1\/4}(1+o(1))$ as  $r\to \iy$.

4) In order to obtain an  analytic extension of the determinant from
the first sheet onto the other sheets we need to use two identities
\er{cS(k)}, \er{cT(k)}.

\medskip

\footnotesize

\no {\bf Acknowledgments.} \footnotesize
A. Badanin was
supported by the RFBR grant  No 16-01-00087.
E. Korotyaev was supported by the RSF grant  No.
15-11-30007.

\end{document}